\documentclass[aps, preprint]{revtex4}
\usepackage{graphicx}
\usepackage{epsf}
\usepackage{wrapfig}
\usepackage{epsfig}
\def\bk{{\mbox{\boldmath$k$}}}
 \newcommand{\ner }{{\bf   }}
\def\ner #1{\mbox{\boldmath$ #1$}}
\def\bq{{\mbox{\boldmath$q$}}}

\def\bp{{\mbox{\boldmath$p$}}}

 \def\br{{\mbox{\boldmath$r$}}}
   
   \def\bP{{\mbox{\boldmath$P$}}}
   \def\bp{{\mbox{\boldmath$p$}}}

\def\br{{\mbox{\boldmath$r$}}}
\def\b0{{\mbox{\boldmath$0$}}}

\def\bk{{\mbox{\boldmath$k$}}}
\def\bq{{\mbox{\boldmath$q$}}}

\def\bp{{\mbox{\boldmath$p$}}}

\def\br{{\mbox{\boldmath$r$}}}
\def\b0{{\mbox{\boldmath$0$}}}

\newcommand{\ra}{\,\rangle}
\newcommand{\la}{\,\langle}

\def \b #1{ {\bf #1}}
\newcommand{\be}{\begin{eqnarray}}
\newcommand{\ee}{\end{eqnarray}}

     \font\tenbifull=cmmib10 scaled 1200 
     \font\tenbimed=cmmib9
     \font\tenbismall=cmmib7
       \def\bmit{\fam9 }
\mathchardef\bbkappa="7114
\mathchardef\bbrho="711A
\mathchardef\bbsigma="711B
\mathchardef\bbtau="711C
\mathchardef\bbvarrho="7125
\mathchardef\bbvarsigma="7126
\mathchardef\bbxi="7118
\def\boldkappa{{\bmit\bbkappa}}

\def\boldtau{{\bmit\bbtau}}

\usepackage{graphicx}
\usepackage{epsf}
\usepackage{wrapfig}
\usepackage{epsfig}
\def\Vec#1{\mbox{\boldmath $#1$}}

\def\beq{\begin{equation}}
\def\eeq{\end{equation}}

\def\beqy{\begin{eqnarray}}
\def\eeqy{\end{eqnarray}}

\newcommand{\ber}{\begin{displaymath}}
\newcommand{\eer}{\end{displaymath}}

\begin{document}
\vskip 2mm
\date{\today}\vskip 2mm
\title{
Slow Proton Production in Semi-Inclusive Deep Inelastic Scattering
 off  Deuteron and Complex Nuclei:
Hadronization and Final State Interaction Effects}

 \author{V. Palli}
 \author{C. Ciofi degli Atti}
  \author{L.P. Kaptari}
  \altaffiliation{On leave from  Bogoliubov Lab.
      Theor. Phys., 141980, JINR,  Dubna, Russia, through the program Rientro dei Cervelli
       of the
      Italian Ministry of University and  Research}
  \author{C. B. Mezzetti}\affiliation{Department of Physics, University of Perugia and
      Istituto Nazionale di Fisica Nucleare, Sezione di Perugia,
      Via A. Pascoli, I-06123, Italy}
      \author{M. Alvioli}\affiliation{104 Davey Lab., The Pennsylvania State University,
      University Park, PA 16803, USA}
\begin{abstract}
\vskip 5mm
\setlength{\baselineskip}{12 pt}
  The effects of the final state interaction  in slow  proton production
  in  semi inclusive deep inelastic
   scattering processes  off nuclei, $A(e,e^\prime p)X$,  are investigated in details
    within  the spectator and target
   fragmentation mechanisms; in the former mechanism, the hard interaction on a nucleon of a
   correlated pair leads, by recoil, to the emission
of the partner nucleon, whereas in the latter mechanism proton is
produced when the diquark, which is formed right after the
$\gamma^*$-quark interaction, captures a quark from the
vacuum. Unlike previous papers on the subject, particular
attention is paid  on the effects of the final state interaction  of the hadronizing
quark with the nuclear medium
   within  an approach based upon
   an effective time-dependent  cross section which combines the soft and hard parts
    of hadronization
  dynamics in terms of the string model and  perturbative QCD, respectively.
 It is shown  that the final state interaction  of the hadronizing quark with the medium plays a relevant role
 both in  deuteron and complex nuclei; nonetheless,
 kinematical regions where  final state interaction  effects are
 minimized can  experimentally be selected, which would allow one to investigate
   the  structure functions of nucleons embedded in the
 nuclear medium; likewise,
 regions where the interaction of the struck hadronizing  quark with the nuclear medium
  is maximized can be found,
  which would make it possible to study non perturbative  hadronization
  mechanisms.
\end{abstract}
\maketitle

\section{Introduction}
Semi Inclusive Deep Inelastic Scattering (SIDIS)  of leptons
\textit{(l)} off nuclei
 can provide relevant
information on: (i) possible modification of the nucleon structure function in
 medium (EMC-like effects),
  (ii) the relevance of exotic configurations at short nucleon-nucleon (NN) distances;
  (iii) the mechanism of quark hadronization.
  A process which attracted much interest from both the theoretical
  (see e.g. \cite{FS}-\cite{marksemi})
  and experimental (see e.g. \cite{experiment}-\cite{wally})
  points of view is the production of slow
  protons,  i.e. the process  A({\it l},{\it l}'p)X, where a slow proton ($p$) is detected
  in coincidence
with the scattered lepton ($\textit{l'}$). In plane wave impulse
approximation (PWIA),
  after the hard collision of the virtual photon $\gamma^*$ with
  a quark of a bound nucleon, two main  production mechanisms of slow protons
  have been considered, namely the {\it spectator} (sp) and the
{\it target fragmentation} (tf) (or {\it direct}) mechanisms. In the former the virtual
photon is assumed to interact with a quark belonging to a nucleon
of a correlated pair: the hit quark leaves the nucleon and
hadronizes, giving rise to a Jet of hadrons, whereas the second correlated nucleon of the
pair recoils with slow momentum and is detected in coincidence
with the scattered lepton; in the target fragmentation mechanism,
slow protons originate from the capture  of  a
quark from the vacuum  by the  spectator diquark. Let us stress that in this paper we do not
consider the production of leading fast protons which arise from
current fragmentation (see e.g. \cite{brooks,leading} for recent experimental advances), although our formalism will be
 generalized to consider this process as well.    In the past,
 several theoretical approaches  to the spectator mechanism have been developed, though
  most of them  either completely disregarded  the final state interaction (FSI),
  or considered only part of it.
  In this paper
    the  results of calculations the cross section within  both the spectator and
   target fragmentation   mechanisms,
    taking  also into  account FSI effects of the hadronizing quark with the nuclear medium
    will be presented. Our paper, which is   motivated by
the results of recent  experiments at JLab \cite{Klimenko}, by our participation
    as theoretical
    support to the  JLab experiment E-03-012 \cite{wally}, and, eventually,
     by the possibility
    to perform SIDIS experiments at the $12\,\, GeV$ upgraded Jlab (see e.g. \cite{brooks}),
    is organized as follows: the general theory of SIDIS is sketched in Section II,
     the SIDIS process on the deuteron and complex nuclei is illustrated in Section III
      and IV, respectively,  and, eventually,
    the Conclusions
    are presented in Section V.
\section{The semi inclusive deep inelastic cross section}
\begin{figure}[!ht]      
\centerline{
\includegraphics[scale=0.4 ,angle=0]{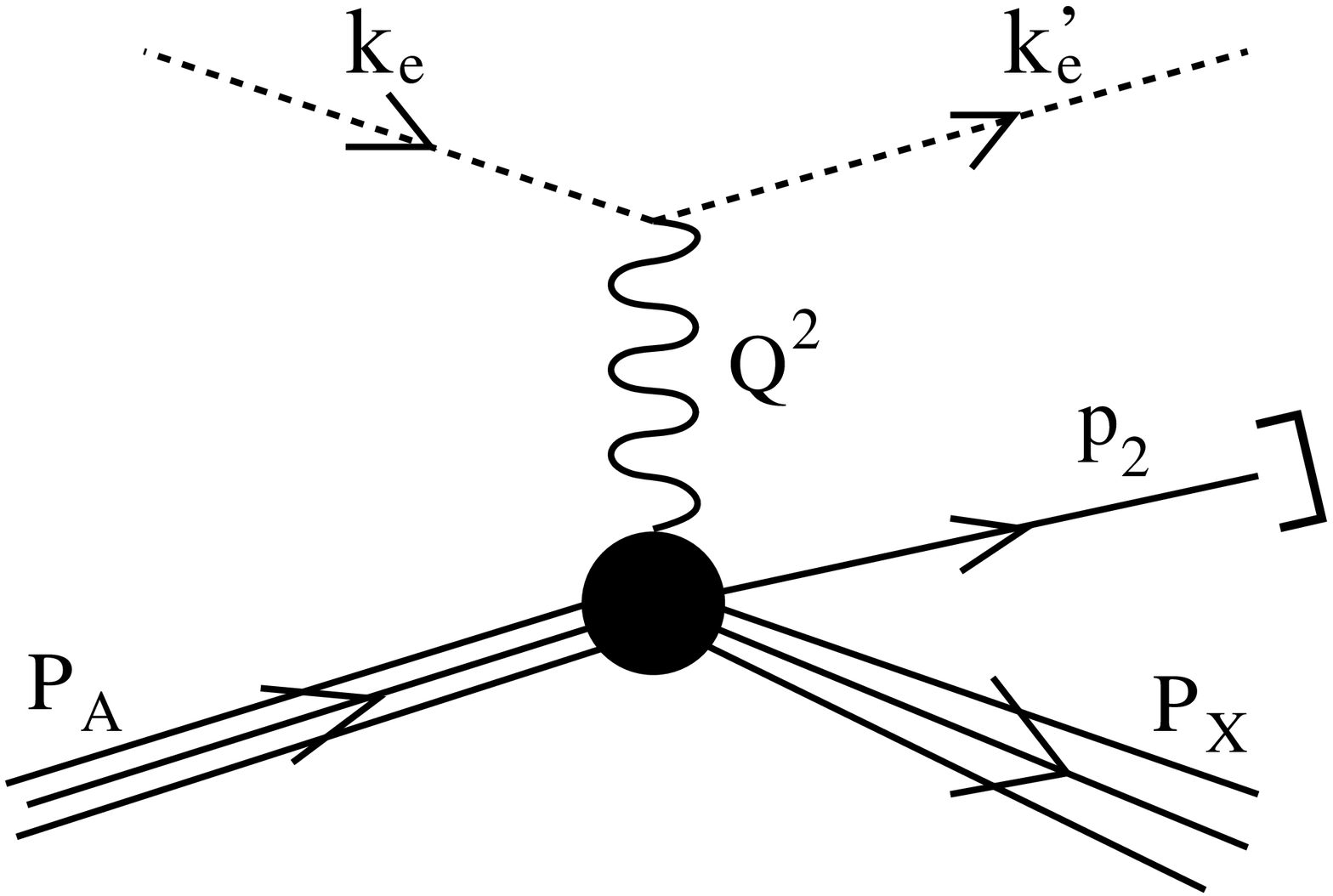}}
\caption{The Feynman diagrams of the process $A(e,e^\prime p)X$
  in one-photon-exchange approximation. The incident electron
  with four-momentum $k_e=(E_e,\bk_e)$ is scattered by the
  nucleus A with four-momentum ${P}_A=(M_A,{\bf 0})$; in the
  final state the scattered electron with four-momentum
  $k_e'=(E_e^\prime,\bk_e^\prime)$ is detected in coincidence
  with a proton with four-momentum
  $p_2=\left (\sqrt{\bp_2^2+m_N^2},\bp_2\,\, \right)$  whereas
  the whole set of undetected particles moves with
  center-of-mass four-momentum $P_X=(E_X,\bP_X)$.
  $Q^2 =-q^2= -(k_e-k_e')^2 = {\bf q}^{\,\,2} - \nu^2=4\,E_e\,E_e^\prime
  sin^2 {\theta_e \over 2}$ is the four-momentum transfer and  $ \theta_e \equiv
  \theta_{\widehat{\bk_e \bk_e^{'}}}$ is  the electron scattering angle.}
  \label{Fig1}
\end{figure}

Within the widely used one-photon exchange approximation, whose
Feynman diagram is shown in Fig.~\ref{Fig1}, the SIDIS cross section
off a nucleus $A$ is given by
\be && \frac{d^4\sigma}{dx dQ^2\
d\bp_2}=\frac{4\alpha_{em}^2}{Q^4}\frac{\pi\nu}{x}
\left[ 1-y-\frac{Q^2}{4E_e^2} \right]
\widetilde{l}^{\mu\nu}L_{\mu\nu}^A = \label{eq1}\\&&
=\frac{4\alpha_{em}^2}{Q^4}\frac{\pi\nu}{x} \left[
1-y-\frac{Q^2}{4 E_e^2} \right] \left [ \tilde l_L W_L
+\tilde l_{T} W_T+\tilde l_{TL} W_{LT}\cos\phi+ \tilde l_{TT}
W_{TT}\cos (2\phi) \right ]. \label{cross} \ee
\noindent Here $\alpha_{em}$ is the fine-structure constant,
 $Q^2 =-q^2= -(k_e-{k_e}')^2 = {\bf q}^{\,\,2} - \nu^2=4 E_e E_e'
sin^2 {\theta_e \over 2}$ the four-momentum transfer, ${\bf q} =
{\bf k}_e - {{\bf  k}_e}'$ and $\nu=E_e -{E_e}' $ the three-momentum
and energy transfer, $ \theta_e \equiv \theta_{\widehat{\bk_e
\bk_e'}}$ the electron scattering angle, $ x = Q^2/2m_N\nu $ the
Bjorken scaling variable,
 $y=\nu /E_e$  and, eventually, $\phi$
is the angle between the scattering and reaction planes.
The four-momentum of the slow detected recoiling nucleon is
denoted by $p_2\equiv(E_2,{\bf p}_2)$ and the Center-of-Mass (CM)
momentum of the whole set of undetected particles  by $P_X\equiv(E_X,{\bf
P}_X)$.
In Eq.
(\ref{eq1})\, $\widetilde {l}_{\mu\nu} $
 and $L_{\mu\nu}^A$ are the electron
 and the nucleus  electromagnetic tensors, respectively;  the former has the well
 known standard
 form, whereas
 the latter can be written as follows
\be &&
 L_{\mu\nu}^A=
  \sum\limits_{X}\la {\bf P}_A|\hat{J}_\mu|{\bf P}_f\ra\la {\bf P}_f|\hat{J}_\nu|{\bf P}_A\ra
(2\pi)^4 \delta^{(4)}\left(k_e+P_A-k_e'-P_X-p_2 \right)
 d\boldtau_X,
\label{eq5}
\ee
where $\hat{J}_\mu$ is the operator of the nucleus
electromagnetic current and  ${\bf P}_A$ and ${\bf P}_f= {\bf
P}_X+{\bf p}_2$ denote the three-momentum of the
target nucleus and the final hadronic state, respectively. Nuclear effects are contained in the
various nuclear responses
  $W_i$, and the quantities $\tilde l_i$ are the
components of the virtual photon
spin density matrix.

As stated in the Introduction, slow proton emission can be due
either to the spectator mechanism or to target fragmentation, the
momentum of the detected nucleon being in both cases small in
magnitude, $p_2\equiv |{\bf p}_2| \lesssim 1\,GeV$,  which is much
less than the value of fast leading hadrons \cite{brooks,leading}
produced in current fragmentation, which therefore  will not be
considered in this paper. It should also be pointed out that both
in the deuteron and complex nuclei cases we consider the momenta
of the detected recoil nucleon always larger than the Fermi
momentum.

\section{Proton production from the deuteron}

 Let us now consider SIDIS of electrons off a  deuteron target, i.e the process of proton production  via the reaction
 \beq
 e+D=e^\prime +p+X,
 \label{Dreac}
 \eeq
 where, we reiterate,  $p$ denotes the produced proton, which is detected in coincidence with the scattered electron, and "X"
 the whole set of undetected particles. This process has
been considered in several theoretical papers \cite{FS}-\cite{marksemi} and experimental investigations
\cite{experiment}-\cite{wally}. Let us first discuss  the spectator mechanism.\\

\subsection{The spectator mechanism}\label{sec:tri}
In the spectator mechanism, depicted in Fig.~2,
  the deep-inelastic electromagnetic  process,
producing the hadronic Jet (${\bf P}_X$=${\bf P}_{Jet}$),
 occurs on the $\it active$ (or "struck")
nucleon, e.g. nucleon "1",  while  the second nucleon  (the  $
spectator$ one)
 recoils with low momentum and is detected in coincidence with the scattered electron.
At high values of the  3-momentum transfer, the Jet (to be also called  "nucleon debris" or
"hadronizing quark")   propagates mainly along the $\bf q$ direction;
within the PWIA (Fig.~2a) it does not interact
 with the slow nucleon, whereas, when the  interaction between the Jet and the spectator nucleon is
 taken into account,  FSI effects
 are generated   (Fig.~2b). The wave function of the final state
 can be written in both cases in the general  form
\be
 && \Psi_f (\{ \xi \},{\bf r}_X,{\bf r}_2 ) =  \
\phi_{\beta_f}(\{ \xi\} )\psi_{{\bf P_X},{\bf p}_2}({\bf r}_X,{\bf r}_2),
\label{eq14}
\ee
 where ${\bf r}_X$ and ${\bf r}_2$  are the
coordinates of the center-of-mass of the Jet $X$ and the
spectator nucleon, respectively, and $\{ \xi\} $ denotes the set of
the internal coordinates of  system $X$; the latter is  described by the internal
wave function $\phi_{\beta_f}(\{ \xi\})$, with $\beta_f$ denoting  all
quantum numbers of the final state, whereas
the wave function  $\psi_{{\bf P}_X,{\bf p}_2}({\bf r}_X,{\bf r}_2)$
describes  the relative motion of  system $X$
and   the spectator nucleon. The matrix elements in  Eq.~(\ref{eq5}) can  easily be  computed, provided
 the  contribution of the two-body  part of the deuteron electromagnetic current
can be disregarded, which means that  the deuteron current  can be
represented as a sum of electromagnetic
 currents of individual nucleons, i.e.
 ${\hat J}_\mu (Q^2,X)={\hat j}_\mu^{N_1} + {\hat j}_\mu^{N_2} $.
Introducing in intermediate states  complete sets of plane waves
 $|{\bf k}_1{'},{\bf k}_2{'} \ra$ and $|{\bf k}_1,{\bf k}_2\ra$, one obtains

\begin{eqnarray}
&&
\la \beta_f,{\bf P}_f={\bf P}_X+{\bf p}_2| {\hat j}_\mu^N| {\bf P}_D\ra=
\nonumber \\&&
\sum\limits_{\beta,{\bf k}_1{'},{\bf k}_2{'}} \sum\limits_{{\bf k}_1,{\bf k}_2}
\la \beta_f,{\bf P}_X,{\bf p}_2|\ \beta, {\bf k}_1',{\bf k}_2' \ra
\la \beta, {\bf k}_1',{\bf k}_2'|j_\mu^{N_1}| {\bf k}_1,{\bf k}_2\ra \la {\bf k}_1,{\bf k}_2| {\bf P}_D\ra =
\nonumber\\&& \int \frac{d^3 k_1}{(2\pi)^3} \psi_D({\bf k}_1)
\la  \beta_f, {\bf k}_1+{\bf q}|{\hat j}_\mu^{N_1} (Q^2,p\cdot q)|{\bf k}_1\ra
\psi_{ {\boldkappa}_f}^+({\bf q}/2+{\bf k}_1),
\label{eq15}
 \end{eqnarray}
 where the matrix element
 $\la \beta_f, {\bf k}_1+{\bf q} |j_\mu^{N_1}(Q^2,k\cdot q)|{\bf k}_1 \ra$
 describes the electromagnetic transition from a moving nucleon
 in the initial  state to the final hadronic system $X$ in a quantum state $\beta_f$.
 Here,  ${\boldkappa}_f=({\bf P}_X-{\bf p}_2)/2$,  and
the  sum over all  final state $\beta_f$ of the
  square of this matrix element, times the corresponding energy conservation
  $\delta$-function, defines  the deep inelastic
  nucleon hadronic tensor for a moving nucleon.
Let us now analyze the PWIA and the FSI cases. For the sake of simplicity we will present our formalism in the target rest frame.

\subsubsection{The PWIA}\label{subsect:triA}

Within the PWIA,  the relative motion of the Jet and the slow proton
is  described by a plane wave
 \be &&
\psi_{\boldkappa_f}({\bf q}/2-{\bf k}_2)\sim (2\pi)^3\delta^{(3)}
({\bf q}/2-{\bf k}_2- \boldkappa_f)= (2\pi)^3\delta^{(3)} ({\bf
k}_2-{\bf p}_2) \label{eq16}
\ee
and the transition matrix element, Eq.(\ref{eq15}), factorizes into the product of
the  matrix element of the  nucleon e.m. current and the deuteron wave function.
As a consequence,
 the four response functions in Eq. (\ref{cross}) can
be expressed in terms of the
 two independent structure functions,  $W_L$ and $W_T$; moreover, if one assumes
 the validity of the
  the Callan-Gross  relation ($2x F_1(x) =  F_2(x)$), the semi inclusive
   cross section (\ref{cross})  depends
   only upon one nucleon  DIS structure function, namely
    $F_2(x)$, i.e.
\be &&
\frac{d^4\sigma_{sp}^{PWIA}}{dx dQ^2\ d\bp_2} = K(x,y,Q^2)\,
n_D(|\bp_2|) \, z_1 F_2^{N_1/D}\left(\frac{x}{z_1}\right),
\label{crossPWIA}
\ee
where $z_1=k_1\cdot q/(m_N\,\nu)$ is the light cone
momentum fraction of the struck nucleon,  and
 the kinematical factor $K(x,y,Q^2)$ is
given by (see, e.g. ref. \cite{scopetta})
\begin{eqnarray}&&
K(x,y,Q^2)=\frac{4\pi\alpha_{em}^2}{Q^4}\frac{1}{x} \left(
\frac{y}{y_1}\right)^2 \left[ \frac{y_1^2}{2} +(1-y_1)
-\frac{k_1^2 x^2 y_1^2}{z_1^2 Q^2} \right ], \label{kinfac}
\end{eqnarray}
where $y_1=\displaystyle\frac{k_1q}{k_1 k_e}$. In the Bjorken limit ($Q^2, \nu
\rightarrow \infty$, $x={\rm const}$)
$y_1=y$ and one has
\beq
K(x,y,Q^2)=\frac{4\pi\alpha_{em}^2}{xQ^4}
\bigg[1-y+\frac{y^2}{2}\bigg].
\label{kfactor}
\eeq
 In Eq. (\ref{crossPWIA}) $F_2^{N/D}({x}/{z_1}) = 2(x/z_1) F_1^{N/D}({x}/{z_1})$ is
  the DIS structure
function of  the  struck ("active") nucleon {\it in the deuteron}
 and $n_D$ is  the momentum
distribution of the struck  nucleon with $|\bk_1|=|\bp_2|$, {\it
viz}
 \be
n_D(|{\bf k}_1|)=\frac13\frac{1}{(2\pi)^3} \sum\limits_{{\cal
M}_D} \left |\int d^3 r
 \Psi_{{1,\cal
M}_D}( {\bf r})\exp(-i{\bf k}_1{\bf  r}/2) \right|^2.
\label{dismom}
\ee
 From what we have exhibited it is clear that if the spectator
mechanism represents the correct description of the process,
it can provide unique information on
the DIS structure function of a nucleon bound in the
deuteron $F_2^{N/D}$ \cite{FS}.
\subsubsection{Final State Interaction (FSI)}
\label{subsect:triB}
The FSI effects account for the reinteraction of the  hadronizing quark
 with the spectator nucleon (Fig.~(2b)). Since the relative motion of the Jet and
 the recoil  proton can no longer be described by a plane wave,
 all four responses contribute, in principle, to the cross section
   ~(\ref{cross}); however a factorization of the nucleon e.m. current and the nuclear
   structure part can be still advocated, provided the following conditions are satisfied
   \cite{ckk,marksemi}: i) $|{\bf q}|$ and $Q^2$ are large enough ($|{\bf q}| \,\geq 1.5 \,\,
   GeV/c$,
    $Q^2 \geq 2.5-5 \,\, (GeV/c)^2$); ii) the rescattering process of the fast system $X$ with the spectator
 nucleon can be considered as an high-energy soft hadronic interaction with small
  momentum transfer in the rescattering process, in which case
   $|\bp_2| \simeq  |{\bf k}_2|$ and
  the matrix element becomes
\be && \la {\bf P}_f| {\hat j}_\mu^N|{\bf P}_D\ra\cong
{\hat j}_{\mu}^N(Q^2,x, {\bf p}_2) \int d^3 r
 \psi_D({\bf r}) \psi_{\boldkappa_f}^+({\bf r})\exp(i{\bf r q}/2).
\label{eq23}
\ee
As a result, the SIDIS cross section can  still be described by
  one structure function $F_2^{N_1/A}$, i.e.
\be
&&
\frac{d^4\sigma_{sp}^{FSI}}{dx dQ^2\ d\bp_2}
=
K(x,y,Q^2)\, n_D^{FSI}({\bf p}_2,{\bf q}) \,z_1  F_2^{N_1/D}\left(\frac{x}{z_1}\right),
\label{crossfsi}
\ee
where
\be
n_D^{FSI}({\bf p}_2,{\bf q})=\frac13\frac{1}{(2\pi)^3} \sum\limits_{{\cal
M}_D} \left |\int d^3 r
 \Psi_{{1,\cal M}_D}( {\bf r}) \psi_{\boldkappa_f}^+({\bf r})\exp(i{\bf r q}/2) \right|^2
\label{dismomfsi1}
\ee
\noindent is the distorted momentum distribution, which coincides with the
 momentum distribution of the hit nucleon
(Eq.~(\ref{dismom}))
 when $\psi_{{\boldkappa}_f}^+({\bf r})\sim \exp(-i{\boldkappa}_f\bf r)$,  with
 $\boldkappa_f={{\bf q}}/{2} - {\bf p}_2 $.
In our  case, when the relative momentum
 is rather large, $\boldkappa_f \sim  {\bf q}/2$, and the rescattering processes occur with low momentum transfers, the
 wave function $\psi_{{\boldkappa}_f}^+({\bf r})$
  can be replaced by its eikonal form describing
  the propagation of the
 nucleon  debris formed after  $\gamma^*$ absorption by a  quark, followed  by
 its hadronization processes and the interaction
 of the newly produced hadrons with the spectator nucleon. This series of soft
 interactions with the spectator can be characterized by an effective cross
 section  $\sigma_{eff}(z,x,Q^2)$ \cite{ciokop}
  depending upon time (or the distance $z$ traveled  by the system $X$).
  Within such a framework,
the distorted  nucleon momentum distribution,  Eq.~(\ref{dismomfsi1}), becomes \cite{ckk}
\begin{equation}
 n_D^{FSI}( {\bf p}_2,{\bf q}) =
\frac13\frac{1}{(2\pi)^3} \sum\limits_{{\cal
M}_D} \left | \int\, d  {\bf r} \Psi_{{1,\cal
M}_D}( {\bf r}) S( {\bf r},{\bf q}) \chi_f^\dag\,\exp (-i
{\bf p}_2 {\bf r}) \right |^2,
 \label{dismomfsi}
\end{equation}
where $\chi_f$ is the  spin function of the spectator nucleon and
$S( {\bf r},{\bf q})$    the $S$-matrix describing
the  final state interaction
between the debris and the  spectator. In Ref. \cite{ckk} the S-matrix has been approximated
by a  Glauber-like eikonal
form, namely
\begin{equation}
S({\bf r},{\bf q}) \equiv G({\bf r},{\bf q})= 1-\theta(z)\,
\Gamma(\ner{b},z),
 \label{G12}
\end{equation}
where   ${\bf r}={\bf r}_1 -{\bf r}_2 \equiv \{{\bf b}, z\}$ and
\begin{equation}
\Gamma(\ner{b},z)\,=\,\frac{(1-i\,\alpha)\,\,
\sigma_{eff}(z)}
{4\,\pi\,b_0^2}\,e^{\displaystyle{-\frac{\ner{b}^{2}}{2\,b_0^2}}}
\label{gamma}
\end{equation}
\noindent is the profile function depending upon $\alpha = Re
f_{NN}(0)/Im f_{NN}(0)$; here  $f_{NN}(0)$ is  the forward elastic
NN scattering amplitude, $\sigma_{eff}$  the effective cross section of interaction
between the
hadronizing quark and the spectator nucleon, and $b_0$
the slope parameter  of the elastic NN scattering amplitude. In Eq. ~(\ref{G12})  the $\theta$
function ensures that rescattering occurs only in the forward hemisphere and
 the dependence upon ${\bf q}$  has been included in order
 to define the orientation of the $z$-axis,
 i.e.
${\bf r}=z\frac
{{\bf q}}{|{\bf q}|} + {\bf b}$, as well as the energy dependence of $\alpha$, $\sigma_{eff}$
 and $b_0$.  Although Eq.~(\ref{G12})
  resembles
the usual Glauber form, it contains an important difference, namely,  unlike the
Glauber case, the
profile function $\Gamma$ depends not only upon the two-nucleon transverse relative separation
 ${\bf b}=
{\bf b}_1-{\bf b}_2$
but also upon the longitudinal separation
$z={z}_1-{z}_2$; this latter dependence is due to the $z$- (or time) dependence of
the effective cross
section $\sigma_{eff}(z)$ obtained in \cite{ciokop}, which describes the interaction of the hadronizing quark, struck
from nucleon
"1", with the spectator nucleon "2".  The effective cross section $\sigma_{eff}(z)$, at the given point $z$,
 consists of   a sum of the  nucleon-nucleon and the  meson-nucleon cross sections
$\sigma_{eff}(z)\,=\,\sigma_{tot}^{NN}\,+\,\sigma_{tot}^{\pi N}\,
\big[\,n_{M}(z)\,+\,n_{G}(z)\,\big]$, where
$n_{M}(z)$ and $n_{G}(z)$ are the effective numbers of mesons produced by the breaking of the
color string and by gluon radiation, respectively.
 As demonstrated  in Ref. \cite{ciokop1},  such an  effective cross section
 provides a good description of grey tracks
 production in  muon-nucleus DIS at high energies \cite{adams}.
 Let us stress  that
 hadronization  is basically a QCD  nonperturbative  process, and, consequently,
 any   experimental information on its  effects on the reaction (\ref{Dreac})
  would be a rather valuable one; since it has been shown in Ref. \cite{ckk} that
   in the kinematical range where the FSI effects
  are relevant the
  process (\ref{Dreac}) is essentially governed by the  hadronization cross section,
 this opens a new and important
  aspect of these reactions, namely  the possibility, through them, to investigate
  hadronization mechanisms
  by choosing a proper kinematics where FSI effects are maximized.

 Let us now consider proton production due to target fragmentation.

\subsection{Target fragmentation}
The target fragmentation (or direct)  mechanism, depicted in Fig.~3 is rather
different from the spectator one; here the detected proton is not
the spectator   proton in the deuteron,  but  a proton which is
formed immediately after the hard ${\gamma^*}$-quark interaction,
when the spectator diquark captures a quark from the vacuum
(note
that in this process ${\bf P}_X$= ${\bf P}_{Jet}$+ ${\bf k}_2$=-${\bf k}_1$).
The cross section corresponding to  the tf mechanism can be calculated by
introducing the notion of nucleon fragmentation function
$H_{1(2)}^{N_1\,N_2}(x,z_2,{\bf p_2}_{\perp})$~\cite{Ffunction}, which
 describes the formation
of nucleon $N_2$ from the hadronization of the diquark of nucleon $N_1$; it is
usually presented  in the following form
 \beq
H_2^{N1,N_2}\left(x,z_2,\ner{p}_{2\bot}^2\right)\,=\,
x\,\rho(\ner{p}_{2\bot})\,\frac{z_2}{1-x}\, \Big[
\sum_q e_q^2f_q(x) D^p_{qq}\left(\frac{z_2}{1-x}\right)\Big]\,,
\label{h2n} \eeq
where  $ z_2=(p_2\cdot q)/m_N\nu \simeq (p_{20}-|{\bf p}_2|\cos \theta_2)/m_N$
is the light cone momentum fraction of the produced proton,
$\rho(\ner{p}_{2\bot})$ is the transverse momentum distribution of
the produced nucleon  with transverse momentum ${\bf p_2}_{\perp}$,
$f_q(x)$  is the parton distribution function,  and, eventually,
$D^p_{qq}(z_2)$ is the diquark fragmentation function representing
the probability to produce a proton with light cone momentum  fraction $z_2$
from a diquark. The explicit
parametrized forms  of  $\rho(\ner{p}_{2\bot})$ and $D^p_{qq}(z_2)$
 can be found, e.g. in Refs.
\cite{distrtr,barframaj}.
 By means of the fragmentation functions, the theoretical analysis
 of  target fragmentation  in SIDIS  becomes
 similar to the theoretical analysis of the spectator mechanisms
 and  a common theoretical framework can be used;
the only difference consists in replacing the deuteron  DIS
structure function $F_{2}^{N/D}(x,{\bf p_2})$ by the deuteron
fragmentation function $H_{2}^{N/D}(x,z_2,\bp_{2\perp}^2)$. Then
in the Bjorken limit the cross section describing the target
fragmentation (tf) mechanism reads as follows
\beq
\frac{d^4\sigma_{tf}}{dx dQ^2\ d\bp_{2}/E_2}
=
K(x,y,Q^2)  H_2^D\left(x,z_2,\ner{p}^2_{2\bot}\right),
\label{tf}
\eeq
\noindent where the kinematical factor $K(x,y,Q^2)$ is given by
  Eq.~(\ref{kfactor}).
The deuteron target fragmentation function
$H_2^D\left(x,z_2,\ner{p}^2_{2\bot}\right)$ can be expressed as a
convolution of the nucleon momentum distributions  and  the
nucleon fragmentation function as follows \beq
H_2^D\left(x,z_2,\ner{p}^2_{2\bot}\right)\,=\,\int\limits_{x+z_p}^{M_D/m_N}
dz_1\,n_D({\bf k}_{1})d^3 {\bf k}_1 \,
\delta\left(z_1-\frac{kq}{m_N\nu}\right)
H_2^{N_1,N_2}\left(\frac{x}{z_1},\frac{z_2}{z_1-x},\left |
\ner{p}_{2\bot} -\frac{z_2}{z_1}{\bf k}_{1\perp}\right|^2\right),
\label{boost} \eeq where   $z_p=z_2(1-x)$ and  and the quantity
$\ner{p}_{2\bot} -\displaystyle\frac{z_2}{z_1}{\bf k}_\perp$ is
the transverse momentum of the detected proton in the rest system
of the struck nucleon
(see e.g. \cite{FS} and  \cite{bosveld}).
Within the considered
kinematics,
 with low and moderate values of the transverse momenta of the
 detected proton, in Eq.~(\ref{boost}) the ${\bf k}_{1\perp}$ dependence is
entirely governed by the momentum distribution $n_D(k_{1z},
 {\bf k}_{1\perp}) \sim exp(-\beta {\bf k}_{1\perp}^2)$ which decreases much faster
($\beta\sim 1.5 fm^2$ for the deuteron and $\beta\sim 3.5-5 fm^2$ for complex
nuclei~\cite{ciosim})   than the nucleon fragmentation function (
$H_2^N\left(x,z,{\bf p}_\perp^2\right)\sim exp(-\beta {\bf p}_{\perp}^2)$
with $\beta\sim 0.38 fm^2$, see below). Then the
transverse  part of the nucleon fragmentation function can be taken
 out of the integral at ${{\bf k}_{1\perp}=0}$
providing
\beq
H_2^D\left(x,z_2,\ner{p}^2_{2\bot}\right)\,\simeq\,\int_{x+z_p}^{M_D/m_N} dz_1\,f_{N_1}(z_1)\,
H_2^{N_1,N_2}\left(\frac{x}{z_1},\frac{z_2}{z_1-x},\ner{p}_{2\bot}^2\right).
\eeq
where
\beq f_{N_1}(z_1)\,=\,2\pi\,m_N\,z_1\,\int_{|{\bf
k}_1^{min}|}^{\infty}\,d|{\bk}_1|\,|{\bk}_1|\,n_D({\bk}_1)\,
\label{fz}
\eeq
is the  light cone momentum distribution of the struck nucleon
   and
$|{\bf k}_1^{min}|=\left
[(m_Nz_1-M_D)^{2\phantom{^1}}-m_N^2]/[2(m_Nz_1-M_D)]\right|$.

\subsection{Numerical results}
  In order to analyze the kinematical conditions under which the effects of FSI
  are minimized
  or maximized, we have considered
  the ratio of the  PWIA cross section
  to the cross section including FSI, given by  Eqs.~(\ref{crossPWIA})  and (\ref{crossfsi}),
  respectively.
We would like to stress here that, whereas in Ref. \cite{ckk} the asymptotic value of
$\sigma_{eff}(z,x)$ \cite{ciokop} has been used, in the present work  we have
obtained the effective cross section at finite values of $Q^2$, $\sigma_{eff}(z,x,Q^2)$, by
the following procedure.
Let us recall
that according to the hadronization model of Ref.~\cite{koppred},
the process of pion production on a nucleon after  $ \gamma^*$
absorption  by a quark can be schematically  represented as in
 Fig.~\ref{Fig4}: at the interaction point  a color string, denoted  $X_1$, and
 a nucleon $N_1$, arising from target fragmentation,
  are formed; the  color  string
propagates  and gluon radiation begins.  The first "pion" is
created  at $z_0 \simeq 0.6$ by the breaking of the color string
and  pion production continues until
it stops at a maximum value of $z =z_{max}$,  when energy conservation does not allow
further
 "pions"
to be created.  We obtain \cite{nashtobe}
\begin{equation}
 z_{max}=\frac {E_{loss}^{max}}{\kappa_{str}+\kappa_{gl}}= \xi
 \frac{E_X -E_N}{\kappa_{str}+\kappa_{gl}}
 \label{nove}
\end{equation}
 after which the number of pions remains constant. Here,
$\kappa_{gl}=2/(3\,\pi) \alpha_{QCD}(Q^2-\Lambda^2)$  ($\Lambda\approx 0.65\,GeV$ and
$\alpha_{QCD}=0.3$) and $\kappa_{str} = 0.2$ represent the energy loss,
 $\kappa = -\frac{dE}{dz}$, of the leading hadronizing quark
 due to the string breaking and gluon radiation, respectively,
 $E_{loss}^{max}= (\kappa_{str} + \kappa_{gl}) z \simeq (E_X- E_{N_1})/2$ is the maximum energy
 loss
 expressed through the energy of the nucleon debris and the energy of the nucleon created by
 target fragmentation at the interaction point.
 Calculation of $z_{max}$  by Eq.~(\ref{nove})
 within the kinematics
  of the
 experiment of Ref. \cite{experiment},  shows that the average number of pions that
 can be created, is about two.
 The results
  of our
  calculations, obtained with the $Q^2$ dependent $\sigma_{eff}(z,x,Q^2)$
   are presented in Fig.~\ref{Fig5}, where
  the angular dependence (left panel) and the dependence
 upon the value of  the spectator momentum (right panel),  are shown at
  $x=0.6$. Kinematics has
  been chosen so as to correspond to the one considered  at the Jlab
  experiments at about  $10 \,\,GeV$. The shaded area reflects the
  uncertainties  in the choice of the parameters appearing in  $\sigma_{eff}$ \cite{ckk}.
   It  can be  seen
  that at low values of momenta and emission in the backward hemisphere,
   the effects of FSI are minimized, so that
  in this region the process $D(e,e^\prime p)X$ could  be successfully used to extract
  the DIS structure function of a bound nucleon.
  Contrarily, at  perpendicular kinematics the FSI effects are rather important and
  essentially depend upon the process of hadronization of the struck quark.
   Therefore, in this region,  the
  processes $D(e,e^\prime p)X$ can serve
  as a source of unique information about
  nonperturbative QCD mechanisms in DIS.
A  systematic experimental study of the processes $D(e,e^\prime
p)X$ is going on at Jlab and first experimental data
at initial electron energy $E_e=5.765 GeV$ are already available
\cite{Klimenko}. At such kinematical conditions the parameters $b_0$ in our calculations
 varies in the range
$0.35-0.6 \,fm$ and $\alpha\simeq -0.35$, with
 a resulting maximum value of $\sigma_{eff}\simeq 100 \,\,mb$.

 In order to minimize the statistical errors, it is common in the
literature to present
 the so called reduced cross section, i.e. the ratio of the
 experimental cross section to all those  kinematical factors,  such that
 in PWIA the theoretical ratio would simply reduce to the product
of  the neutron DIS structure function $F_2^n(x,Q^2)$ times the
deuteron momentum distribution (\ref{dismom}). Thus any
deviation from such a product
should be ascribed to  the failure of the PWIA, due  either to  deviations of
the free structure function from the bound one, or to FSI effects.
In Fig.~\ref{Fig6} the experimental reduced cross section
\cite{Klimenko} is compared with our theoretical results obtained
within the spectator mechanism  in PWIA (dashed curves) and taking
FSI into account (full curves). It can be seen that: i)
 the spectator mechanism within the PWIA does not explain the data in
the whole kinematical range, and ii)  the inclusion of the FSI
between the hadronizing quark and the spectator appears to be
necessary to explain the data. We would like to point out that the reduced cross
section is generated by the interplay between the PWIA and FSI.
At low values of $|{\bf p}_2| \simeq 0.2-0.3\,\,GeV/c$, the interference between PWIA and FSI mostly
cancels out, whereas at high values of  $|{\bf p}_2|$ the deuteron wave function drops out very fast
and, at perpendicular kinematics, the reduced cross section is dominated by eikonal-type FSI.
The fact that the calculated reduced cross section at large values of $|{\bf p}_2|$
appears to agree with the experimental data make us confident that our approach to FSI
is basically correct.
In closing our analysis of the
spectator mechanism, we would like to point out that,   besides
our previous work \cite{ckk} and the present paper, FSI between the nucleon
debris and the spectator nucleon, has also been taken into account
in Ref. \cite{marksemi} by an approach in which  the scattering amplitude
describing the rescattering  between the debris and the
spectator nucleon has been chosen in the form
$f=\sigma_{eff}(i+\alpha)exp(-\frac{1}{2}B^2\,k_{\perp}^2)$ with
$\alpha$, $B$ and $\sigma_{eff}$ as free parameters; in particular
$\sigma_{eff}$ has been varied  in the range $ 0-80\,\,mb$, and $B$ and $\alpha$ have been fixed at
 $B= 8\,\, GeV^2$ and
$\alpha=-0.2$;  the
effects of FSI appear to be in qualitative agreement with our
results, which can be understood  in light of the fact that,
according to our hadronization model, only  two pions can be
produced in the kinematics of Ref. \cite{experiment}.

In order to
estimate the role of the target fragmentation mechanism, we have
calculated the  ratio
\beq
R =\,\frac{d\sigma_{tf}+d\sigma_{sp}^{PWIA}}{d\sigma_{sp}^{PWIA}}\,,
\eeq
 which, obviously,  characterizes the relative contribution
of the fragmentation cross section. The transverse hadron momentum
distribution appearing in Eq. (\ref{h2n}) has been parametrized in the following form \cite{distrtr}
\beq
\rho(\ner{p}_{2\bot}) = \frac{\beta}{\pi}\,
\exp(-\beta\ner{p}_{2\bot}^2)\,,
\eeq
with  $\beta=<\ner{p}_{2\bot}^2>^{-1} = 0.38\,fm^{2}$,  while the fragmentation function
 $D_{qq}$  has been  taken from
Ref.~\cite{barframaj},
both choices being fully satisfactory for the purpose of the
present paper. The results of  calculations are shown in
Fig.~\ref{Fig7}, where $R_{tf}$ is presented {\it vs.} the
emission angle of the detected proton at several fixed values of
the momentum (left Figure),  and {\it vs.} the spectator momentum at  fixed
emission angles (right Figure). As expected, the fragmentation
mechanism contributes only in a very narrow forward direction and
for large values of the spectator momentum.
We would like to stress that the ratio between the direct (target fragmentation)
and spectator
cross sections of
the process $D(e,e'p)X$ has been analyzed in detail in Ref. \cite{FS}
 within the light front (LF) dynamics,  {\it vs.} $x$ and  ${\bf p}_{2\perp}=0$,
  using LF deuteron wave functions corresponding to
 the RSC interaction  (cf  Fig. 3.8 of Ref. \cite{FS}).
  Our results shown in Fig. \ref{Fig7} are in good  agreement with the ones in Ref. \cite{FS}.
\section{complex nuclei}
\subsection{The spectator mechanism }

Let us first of all point out that in a complex nucleus the spectator
mechanism can only occur on a correlated nucleon-nucleon pair,
for
if $\gamma^*$ interacts with a mean field nucleon, the most
probable event would be the coherent recoil of the $(A-1)$-nucleon
system. In order to describe the spspectator mechanism one needs therefore a model
of nucleon-nucleon (NN) correlations in nuclei.
In the so called  strict two-nucleon correlation (2NC) model
the whole nucleus momentum ($\sum _{i=1}^A {\bf k}_i=0$) is shared
by  two correlated nucleons, with  equal and opposite momenta,
with the $(A-2)$-nucleon system at rest, i.e.  ${\bf K}_{A-2}=0$.
On the contrary, in the  few-nucleon correlation (FNC) model a small part of
the momentum is also carried out by the $(A-2)$-nucleon system,
i.e. ${\bf K}_{A-2}= -{\bf k}_{cm} \neq 0$,  ${\bf
k}_{cm}$ being the center-of-mass (CM)  momentum of the correlated
pair. Thus if $\gamma^*$ interacts with one correlated nucleon of
the pair, the partner nucleon recoils and is detected. The process
is similar to the one on a free deuteron, the main difference
being the CM motion of the pair and different types of FSI which occur in a complex nucleus.
 In this section  our approach is generalized to complex nuclei
in the same way as it has been done in Ref. \cite{ciosim2}, with
the relevant difference that in the present paper also the FSI of
the hadronizing quark with the spectator  nucleons is taken into
account. We start with the PWIA and then will consider the effects of the FSI.

\subsubsection{The PWIA}
As already pointed out,  the spspectator mechanism in
complex nuclei can occur only on a correlated nucleon pair, since
in the independent particle model without correlations the whole
system $(A-1)$ would recoil.
Thus, in PWIA, the cross section of the process we are considering has to be
proportional to the joint probability to find in the ground state of the target nucleus
  two correlated nucleons with momenta
 $\bk_1$ and $\bk_2$ and removal energy $E^{(2)}$; this quantity
 is nothing but the well known two-nucleon correlated spectral function, i.e.
 the following quantity
 \beqy
 P_{N_1,N_2}(\bk_1,\bk_2,E^{(2)})&=&\langle \Psi_A^0|a^{\dag}_{\bk_2}a^{\dag}_{\bk_1}\delta
 (E^{(2)}-(H_{A-2}-E_A))a_{\bk_1}a_{\bk_2}|\Psi_A^0\rangle\nonumber\\
 &=&\sum_f\left|\langle\Phi_{\bk_1,\bk_2}, \Psi^f_{A-2}|\Psi_A^0 \rangle \right|^2
  \delta(E^{(2)}-(E^f_{A-2}-E_A)),
 \label{spectral2}
 \eeqy
where ${a^{\dag}_{\bk}} ({a_{\bk}})$ are nucleon creation (annihilation)
operators, $\Psi_A^0$ is the ground state wave function of the
target, eigenfunction of the Hamiltonian $H_A$ with (positive)
eigenvalue $E_A$, $\Psi^f_{A-2}$ is the eigenfunction of the
Hamiltonian $H_{A-2}$ with (positive) eigenvalue
$E^f_{A-2}=E_{A-2}+E^*_{A-2}$=$E_{A-2}+E^{(2)}-E_{thr}^{(2)}$,
where $E_{A-2}$ is the (positive) ground-state energy of the
 $(A-2)$ nucleus  and
$E_{thr}^{(2)}$=$2m_N+M_{A-2}-M_A$ is the two-nucleon threshold
energy.
Because of the lack of  realistic many-body two-nucleon
spectral functions  for finite nuclei, and also given the exploratory
nature of the present work,
we will use here, as in Ref. \cite{ciosim2}, the two-nucleon spectral
function resulting from the FNC model; in this model the two-nucleon spectral function
coincides with the decay function introduced in Ref. \cite{FS} and
represents the probability that, after a nucleon with momentum
$\bk_1$ is instantaneously removed from the target, the residual
$(A-1)$-nucleon system  decays into
 a nucleon with momentum $\bk_2$ and an $(A-2)$-nucleon system in the ground
 or in a well defined energy state (in this respect, the process we are considering is
 a semi-exclusive process rather than a semi-inclusive one; we will come back to
 this point later on).
  The FNC model spectral function
 (\ref{spectral2}) for the deuteron is simply the momentum distribution,
 whereas for $^3He$ is the three-body wave function in momentum space, times the corresponding energy
 delta function.
 For a generic nucleus with $A>3$  one has
 \cite{ciosim}
 \beqy
\label{2ncex}
P_{N_1,N_2}(\Vec{k}_{1},\Vec{k}_{2},E^{(2)})&=&
\frac{n^{rel}_{N_1,N_2}(|\Vec{k}_{1}-\Vec{k}_{2}|/2)}{4 \pi}
\,\frac{n^{cm}_{N_1,N_2}(|\Vec{k}_{1}+\Vec{k}_{2}|)}{4 \pi}\,\delta(E^{(2)}-E_{th}^{(2)})
\eeqy
which, using momentum conservation $\Vec{k}_{1}+\Vec{k}_{2}=-\Vec{K}_{A-2}= \Vec{k}_{cm}$, can also be written as follows
\beqy
P_{N_1,N_2}(\Vec{k}_{cm}/2-\Vec{k}_{2},\Vec{k}_{2},E^{(2)})=
\frac{n^{rel}_{N_1,N_2}(|\Vec{k}_{cm}/2-\Vec{k}_{2}|)}{4 \pi}
\,\frac{n^{cm}_{N_1,N_2}(|\Vec{k}_{cm}|)}{4 \pi}\,\delta(E^{(2)}-E_{th}^{(2)}),
\label{spfun}
\eeqy
\noindent where, in both equations, $n^{rel}_{N_1,N_2}$ and $n^{cm}_{N_1,N_2}$ are the relative and
center-of-mass momentum distributions of the correlated pair $(N_1,N_2)$.

The calculation of the PWIA diagram  of Fig. \ref{Fig8}(a)  yields
\beq
\frac{d\sigma_{sp}^{PWIA}}{dxdQ^2d\bp_2}\,=\,K(x,y,Q^2)\,F_2^{N_1/A}(x,\ner{p}_2),
\label{compl}
\eeq
 with the factor $K(x,y,Q^2)$ given by Eq. (\ref{kfactor}) and the SIDIS
  nuclear structure function $F_2^{N_1/A}(x,\bp_2)$  defined as follows \cite{ciosim2}
\beqy
F_2^{N_1/A}(x,{\bf p}_2)&=&m_N\sum_{N_2}\int_x^{M_A/m_N-z_2}
dz_1\,z_1\,F_2^{N}\left(\frac{x}{z_1}\right)\times \nonumber\\
&\times&\int
d\Vec{k}_{cm}\,
\frac{n^{rel}_{N_1,N_2}(|{\bf k}_{cm}/2-\Vec{p}_{2}|)}{4
\pi} \,\frac{n^{cm}_{N_1,N_2}(|\Vec{k}^{cm}|)}{4 \pi}
\times \nonumber\\
&\times& \delta(M_A-m_N(z_1+z_2)-M_{A-2}z_{A-2}),
\label{qualesara}
\eeqy
where  $\Vec{k}_{cm}=\Vec{k}_1 + \Vec{k}_2=-{\bf P}_{A-2}$ and $\Vec{k}_{2}=\Vec{p}_2$. Here
$F_2^N({x}/{z_1})$ is the structure function of
the hit nucleon, and
 $z_2=[(m_N^2+{\bf p}_2^2)^{1/2}  - |{\bf p}_2|\cos\theta_2]/m_N$, and
$z_{A-2}=[( (M_{A-2})^2+{\bf k}_{cm}^2 )^{1/2} + {\bf k}_{cm}\cdot\bq/|\bq| ]/M_{A-2}$ are
 the light-cone momentum fractions of the  detected nucleon and the recoiling
 spectator nucleus $(A-2)$, respectively.

\subsubsection{The FSI}

The treatment of the FSI  in complex nuclei is more involved than in the deuteron
 since, as already pointed out,   the structure of the Spectral
Function (\ref{spfun})   implies that $(A-2)$ is in the ground or in a well
defined energy state;
 in this case, after $\gamma^*$ absorption,  the final state consists
 of at least three different interacting  systems
 (c.f. Fig.~\ref{Fig8}(b)):
the undetected hadron debris $X$, the undetected $(A-2)$-nucleon system
and, eventually, the detected proton $p_2$. Correspondingly, the
FSI can formally be divided into three classes~\cite{noi}, namely:
i) the FSI of the hadron debris with the spectator
$(A-2)$-nucleon system; ii) the interaction of the
 recoiling nucleon  with the $(A-2)$-nucleon system;
 iii) the interaction of  the  hadron debris with the recoiling proton. Note that
  FSI of the type i) reduces the survival probability
 of having $(A-2)$ in the ground state,  and that of the type ii) and iii)  reduce
  the survival probability of the
 struck proton.  Note, moreover,
  that in the spectator mechanism one has ${\bf P}_X={\bf P}_{Jet}+ {\bf P}_{A-2}$, whereas in the
  target fragmentation process one has ${\bf P}_X={\bf P}_{Jet}+ {\bf P}_{A-1}$.
       \\

\noindent (a) {\it The distorted Spectral Function}\\

The FSI of the hadronizing quark with the $(A-2)$-nucleon system and with the spectator
nucleon, is treated in the
same way as in the deuteron case, i.e.  by using
the effective cross section $\sigma_{eff}$ within the eikonal approximation.
Then in Eq. (\ref{qualesara}) the spectral
function $P_{N_1,N_2}(\bk_1,\bk_2,E^{(2)})$
 has to be replaced with the \textit{Distorted
spectral function}, which can be written in the following way
\beqy
P^{FSI}_{N_1,N_2}(\bk_{1},\bp_{2},E^{(2)})&=&
\sum_{f}\,|T_{fi}|^{2}\,\delta(E^{(2)}-E_{th}^{(2)})=\nonumber\\
&=&\sum_{f}\,\left|\langle\bP_{Jet},\bp_{2},\Psi_{A-2}^{f}(\bk_{3},...,\bk_{A}),
\hat{S}_{FSI}\,|\,\bq,\Psi_{A}^{0}(\bk_{1},\bk_{2},...,\bk_{A})\rangle\right|^{2}\times \nonumber\\
&\times&\delta(E^{(2)}-E_{th}^{(2)}),
\eeqy
where $\hat{S}_{FSI}$ is the FSI operator and $T_{fi}$  the transition matrix element of
the process
having the following form
\beqy
T_{fi}&=&\frac{1}{(2\pi)^6}\,\int \prod_{i=1}^{A}d\br_{i}\,
e^{-i\bP_{Jet}\cdot\br_{1}}\,e^{i\bq\cdot\br_{1}}
\,e^{-i\bp_{2}\cdot\br_{2}}\,\times\nonumber\\
&&\hspace{2cm}\times\,\Psi_{A-2}^{\dag f}(\br_{3},
\dots,\br_{A})\,\hat{S}_{FSI}(\br_{1},\dots,\br_{A})
\,\Psi_{A}^{0}(\br_{1},\dots,\br_{A})\,.
\label{tfi1}
\eeqy
According to our classification of the FSI effects, the operator $\hat{S}_{FSI}$
will read as follows
\beq \label{fsiop}
\hat{S}_{FSI}(\ner{r}_{1},\ner{r}_{2},\dots,\ner{r}_{A})\,=\,D_{{\bp}_{2}}(\ner{r}_{2})
\,G(\ner{r}_1,\ner{r}_2)\prod_{i=3}^{A}G(\ner{r}_1,\ner{r}_i)\,,
\eeq
where $D_{{\bp}_{2}}(\ner{r}_{2})$ and
$G(\ner{r}_1,\ner{r}_2)$ take care, respectively, of  the interaction of the
slow recoiling proton with $(A-2)$-nucleon system and with the
fast nucleon debris, whereas
$\prod_{i=3}^{A}G(\ner{r}_1,\ner{r}_i)$ takes into account the
interaction of the latter with $(A-2)$-nucleon system. Properly generalizing  our
previous treatment of the deuteron case, we have
\beq
\prod_{i=2}^A\,G(\ner{r}_1,\ner{r}_i)\,=\,\prod_{i=2}^A\,\Big[1-\theta(z_{i}-z_{1})\,
\Gamma(\ner{b}_{1}-\ner{b}_{i},z_{i}-z_{1})\Big],
\eeq
 where
$\ner{b}_i$ and $z_i$ are the transverse and longitudinal
components of the coordinates of nucleon ``i'', and the function
$\theta(z_{i}-z_{1})$ describes forward debris propagation, and
$\Gamma$ is given by Eq. (\ref{gamma}).

As far as the  FSI of the  recoiling nucleon with the residual nucleus
$(A-2)$-nucleon system is
concerned, following Ref. \cite{ciosim2}, we have treated it by
   an Optical Potential approach, according to which
the outgoing nucleon plane wave is distorted by
the eikonal phase factor
\beq
e^{-i\,\ner{p}_{2}\cdot\ner{r}_{2}}\longrightarrow
e^{-i\,\ner{p}_{2}\cdot\ner{r}_{2}} D_{\bp_{2}}(\ner{r}_{2}), \eeq
where
\beq
D_{\bp_{2}}(\ner{r}_{2})\,=\,exp\left(-i\,\frac{E_2}{\hbar
|\bp_{2}|}\, \int_{z_{2}}^{\infty}dz\,V({\bf b}_2,z)\right)\,.
\label{Dp2}
\eeq
 We used an energy dependent complex optical potential with
the real and imaginary parts given, respectively, by
\beq Re
V({\bf r})\,=\,-\frac{\hbar \,|{\bf p}_2|}{E_2}\, \frac {\alpha
\,\sigma_{tot}^{NN} \rho ({\bf r})} {2} \label{real} \eeq
and
\beq
Im V({\bf r})\,=\,-\frac{\hbar \,|{\bf p}_2|}{E_2}\,\frac
{\sigma_{tot}^{NN} \,\rho ({\bf r})} {2} \label{im},
\eeq
where
$\rho$ is the one-body  density and $\sigma_{tot}^{NN}$  the total
NN cross section. When the energy of the propagating proton is
small, each rescattering causes a considerable loss of
energy-momentum and the flux of the outgoing proton plane
 wave is
suppressed by the imaginary part of the potential.

Using in Eq. (\ref{tfi1})  momentum conservation
$\ner{P}_{Jet}=\ner{q}-\ner{p}_{2}-\ner{P}_{A-2}$, the transition
matrix element of the process $A(e, e^\prime p)X$ becomes:
\beqy
T_{fi}&=&\frac{1}{(2\pi)^6}\int \prod_{i=1}^{A}d\ner{r}_{i}\,
e^{i(\ner{P}_{A-2}+\ner{p}_{2})\cdot\ner{r}_{1}}
e^{-i\ner{p}_{2}\cdot\ner{r}_{2}}\times\nonumber\\
&&\times\,\Psi_{A-2}^{\dag f}(\ner{r}_{3},
\dots,\ner{r}_{A})\hat{S}_{FSI}(\ner{r}_{1},\dots,\ner{r}_{A})
\Psi_{A}^{0}(\ner{r}_{1},\dots,\ner{r}_{A})\,=\nonumber\\
&=&\frac{1}{(2\pi)^6}\int d\ner{r}_{1}d\ner{r}_{2}
\,e^{i(\ner{P}_{A-2}+\ner{p}_{2})\cdot\ner{r}_{1}}e^{-i\ner{p}_{2}\cdot\ner{r}_{2}}
\,I^{FSI}(\ner{r}_{1}, \ner{r}_{2})\,,
\eeqy
where
\beq
I^{FSI}(\ner{r}_{1}, \ner{r}_{2})\,=\,\int \prod_{i=3}^{A}d\ner{r}_{i}\,
\Psi_{A-2}^{\dag f}(\ner{r}_{3},\dots,\ner{r}_{A})
\hat{S}_{FSI}(\ner{r}_{1},\dots,\ner{r}_{A})
\Psi_{A}^{0}(\ner{r}_{1},\dots,\ner{r}_{A})
\label{twobodyphi}
\eeq
is the distorted two body overlap integral.

 We reiterate,  that in the present approach we consider protons
 with relatively large momenta (at the average  Fermi momentum scale)
 originating  from  correlated pairs in the parent nucleus.
Then for such kinematics the nuclear wave function  can be written
as follows~\cite{ciosim}
 \beq
\Psi_{A}^{0}(\ner{r}_{1},\dots,\ner{r}_{A}) \,=\,
\sum_{\alpha\beta} \Phi_{\alpha}(\ner{r}_1,
\ner{r}_2)\otimes\Psi_{A-2}^{\beta}(\ner{r}_{3},\dots,\ner{r}_{A})\,,
\label{2NCwf}
\eeq
 where $\Phi_{\alpha}(\ner{r}_1, \ner{r}_2)$ and
$\Psi_{A-2}^{\beta}(\ner{r}_{3},\dots, \ner{r}_{A})$ describe the
correlated pair and the $(A-2)$-nucleon system remnants,
respectively. In Eq.~(\ref{2NCwf}) the symbol $\otimes$ is used
for a short-hand notation of the corresponding Clebsh-Gordon
coefficients. The wave function of the correlated pair can be
expanded over a complete set of wave functions describing the
intrinsic state of the pair and its motion   relative to the
$(A-2)$-nucleon system kernel, viz.
\beq
\Phi_{\alpha}(\ner{r}_{1},
\ner{r}_{2})\,=\,\sum_{mn}\,c_{mn}\,\phi_m(\ner{r})\chi_n(\ner{R})\,,
\eeq
where $\ner{R}=\frac{1}{2}(\ner{r}_1+\ner{r}_2)$ and
$\ner{r}=\ner{r}_1-\ner{r}_2$ are the center of mass and relative
coordinate of the pair. As already mentioned, in the FNC  model it
is assumed that the correlated pair carries   most part of the
nuclear momentum, while the
 momentum of the relative motion of the pair and $(A-2)$ nucleus is small~\cite{ciosim}.
 This allows one to treat the  CM motion   in its   lowest $^1S_0$ quantum state
 (in what follows denoted, for the sake of
 brevity, as $os$-state).
We can therefore write
\beq
\Phi_{\alpha}(\ner{r}_{1},
\ner{r}_{2})\,\simeq\,\chi_{os}(\ner{R})\,\sum_{m}c_{mo}
\,\phi_m(\ner{r})\,=\,
\chi_{os}(\ner{R})\,\varphi(\ner{r}) \label{f12}
\eeq
 with
 \beq
\varphi(\ner{r})\,=\,\sum_{m}\,c_{mo}\,\phi_m(\ner{r}).
\eeq
Finally we have
 \beq
\Psi_{A}^{0}(\ner{r}_{1},\dots,\ner{r}_{A})\,\simeq\,\chi_{os}(\ner{R})\,\varphi(\ner{r})
\,\Psi_{A-2}^{0}(\ner{r}_{3},\dots,\ner{r}_{A})\,.
\eeq
Placing
this expression in Eq. (\ref{twobodyphi}) we get:
\beq
I^{FSI}(\ner{r}_1,\ner{r}_2)\,=\,\int
\prod_{i=3}^{A}d\ner{r}_{i}\,
\chi_{os}(\ner{R})\,\varphi(\ner{r})\,\hat{S}_{FSI}(\ner{r}_{1},\dots,\ner{r}_{A})
\,|\Psi_{A-2}^{0}(\ner{r}_{3},\dots,\ner{r}_{A})|^2
\eeq
 and,
disregarding correlations in the $(A-2)$-nucleon system, one can
write \cite{glau1,glau2}:
 \beq
|\Psi_{A-2}^{0}(\ner{r}_{3},\dots,\ner{r}_{A})|^2\,\simeq\,\prod_{i=3}^A\rho(\ner{r}_i)\,,
\eeq
with $\int\rho(\ner{r}_i)d\ner{r}_i=1$, so that, eventually,
the distorted overlap integral becomes:
\beqy
I^{FSI}(\ner{r}_1,\ner{r}_2)&=&\int\prod_{i=3}^{A}d\ner{r}_{i}\,
\phi(\ner{r}_{1}, \ner{r}_{2})\prod_{i=3}^A\rho(\ner{r}_i)
D_{p_{2}}(\ner{r}_{2})\,G(\ner{r}_1,\ner{r}_2)\prod_{i=3}^{A}G(\ner{r}_1,\ner{r}_i)=
\nonumber\\
&=&\phi(\ner{r}_1, \ner{r}_2)\,G(\ner{r}_1,\ner{r}_2)
\,D_{p_{2}}(\ner{r}_{2})
\bigg[\int d\ner{r}\,\rho(\ner{r})\,G(\ner{r}_1,\ner{r})\bigg]^{A-2},
\eeqy
where $\phi(\ner{r}_{1}, \ner{r}_{2})\,=\,\chi_{os}(\ner{R})\,\varphi(\ner{r})$
(cf. Eq. (\ref{f12})).
 In our calculations
the function $\phi(\ner{r}_{1}, \ner{r}_{2})$ has been chosen
 in
such a way, that in PWIA  the same high momentum components
 of the two-nucleon spectral function, as reported in Ref.~\cite{ciosim} are obtained.
Disregarding the real
part of the forward scattering amplitude and considering $A>>1$
  we can write:
\beqy
\hspace{-0.5cm}\bigg[\int d\ner{r}\,\rho(\ner{r})G(\ner{r}_1,\ner{r})\bigg]^{A-2}
\hspace{-0.5cm}&=&
\bigg[\int d\ner{r}\,\rho(\ner{r})-\int d\ner{b}\,\int^\infty_{z_1}dz\,\rho(\ner{b},z)
\,\Gamma(\ner{b}_1-\ner{b};z-z_1)\bigg]^{A-2}\simeq\nonumber\\
&\simeq&\bigg[1-\frac{1}{2}\int_{z_{1}}^{\infty} dz\,\rho(\ner{b}_1,z)
  \,\sigma_{eff}(z-z_1)\bigg]^{A-2}\,\simeq\nonumber\\
&\simeq&exp\left(-\frac{1}{2}\,A\int_{z_{1}}^{\infty} dz\,\rho({\bf b}_1,z)
  \,\sigma_{eff}(z-z_1)\right)
\eeqy
which represents the probability that the debris and the proton did not interact.
Finally we can write the transition matrix element in the following way:
\beqy
T_{fi}&=&\frac{1}{(2\pi)^6}\int d\ner{r}_{1}\,d\ner{r}_{2}\,
e^{i\,\left(\ner{P}_{A-2}\,+\,\ner{p}_{2}\right)\cdot\ner{r}_{1}}\,e^{-i\,\ner{p}_{2}\cdot\ner{r}_{2}}
\,\phi(\ner{r}_1, \ner{r}_2)\,\times\nonumber\\
&&\times\,G(\ner{r}_1,\ner{r}_2)\,D_{p_2}(\ner{r}_2)
\,exp\left(-\frac{1}{2}\,A\int_{z_{1}}^{\infty} dz\,\rho({\bf b}_1,z)\,\sigma_{eff}(z-z_1)
\right)
\label{tfi}
\eeqy
and the Distorted Spectral Function is eventually
\beqy
P^{FSI}_{N_1,N_2}(-(\bP_{A-2}+{\bf p}_2),\bp_{2},E^{(2)})&=& \left|\frac{1}{(2\pi)^6}\int d\ner{r}_{1}\,d
\ner{r}_{2}\,
e^{i\,\left(\ner{P}_{A-2}\,+\,\ner{p}_{2}\right)\cdot\ner{r}_{1}}\,e^{-i\,\ner{p}_{2}\cdot\ner{r}_{2}}
\,\phi(\ner{r}_1, \ner{r}_2)\,\times\right.\nonumber\\
&\times&\left.\,G(\ner{r}_1,\ner{r}_2)\,D_{{\bf p}_2}(\ner{r}_2)
\,\exp \left[-\frac{1}{2}\,A\int_{z_{1}}^{\infty} dz\,\rho(\ner{b}_1,z)\,\sigma_{eff}(z-z_1)
\right]\right|^2\times \nonumber\\
&\times&\delta(E^{(2)}-E_{th}^{(2)}).
\label{Pfi}
\eeqy
which reduces to  the usual Spectral Function  (Eq. (\ref{spectral2}),
 with ${\bf k}_{cm}= -{\bf P}_{A-2}$)  in absence of any FSI.
The sm  cross section becomes
\beq
\frac{d^4\sigma^{FSI}_{sm}}{dxdQ^2d\bp_2}\,=\,K(x,y,Q^2)\,F_2^{(N_1/A,\,FSI)}(x,\ner{p}_2),
\label{sigmafsi}
\eeq
 with the factor $K(x,y,Q^2)$ given by Eq. (\ref{kfactor}) and
 the SIDIS nuclear structure function $F_2^{(N_1/A,\,FSI)}(x,\ner{p}_2)$
 being
\beqy
F_2^{(N_1/A,\,FSI)}(x,\ner{p}_2)&=&m_N\sum_{N_2}\int_x^{M_A/m_N-z_2}
dz_1\,z_1\,F_2^{N}\left(\frac{x}{z_1}\right)\times \nonumber\\
&\times&\int d\Vec{P}_{A-2}\,dE^{(2)} P^{FSI}_{N_1,N_2}(-({\bf
P}_{A-2}+{\bf p}_2),{\bf p}_2,E^{(2)})
\times \nonumber\\
&\times& \delta(M_A-m_N(z_1+z_2)-M_{A-2}z_{A-2})
\label{crosfsi}
\eeqy
with $P_{N_1,N_2}^{FSI}$ given by Eq. (\ref{Pfi}).
It can be seen that  in absence of any FSI, the PWIA results,
given by Eq. (\ref{qualesara}), is recovered. \\

\subsection{The target fragmentation mechanism}
\noindent
Let us now consider  proton production  from the target fragmentation mechanism, in which the
quark-gluon debris originates from current fragmentation,  and the proton
from target fragmentation
 (cf. Fig.~\ref{Fig8}(c)).
 The corresponding cross section can be expressed in terms of  two nuclear
structure functions $H_1^A$ and $H_2^A$ as follows
\beq
\frac{d^4\sigma_{tf}}{dx\,dQ^2\,d \bp_2/E_2} =
\frac{4\pi\alpha^2 }{xQ^4}\,\Bigg[x\,y^2\,H_1^A\left(x, z_2, \ner{p}_{2\bot}^2\right)
\,+\,(1-y)\,H_2^A\left(x,z_2, \ner{p}_{2\bot}^2\right)\Bigg]\,,
\eeq
where  $H_{1(2)}^A$ can be written  as a
convolution of the nucleon fragmentation  function and the
nuclear spectral function of nucleon "1", $P_{N_1}(|\bk_1|,E)$, as follows
\beq
H_1^A(x,z_2,\ner{p}_{2\bot}^2)\,=\,\int dz_1\,f_{N_1}(z_1)\,\frac{1}{z_1}\,
H_1^{N_1,N_2}\left(\frac{x}{z_1},\frac{z_2}{z_1-x},\ner{p}_{2\bot}^2\right)\,,
\eeq
\beq
H_2^A\left(x,z_2,\ner{p}_{2\bot}^2\right)\,=\,\int dz_1\,f_{N_1}(z_1)\,
H_2^{N_1,N_2}\left(\frac{x}{z_1},\frac{z_2}{z_1-x},\ner{p}_{2\bot}^2\right)\,,
\eeq
where $H_{1}^{N_1,N_2}$ and $H_{2}^{N_1,N_2}$ are the fragmentation structure functions of the
struck nucleon $N_1$ producing the detected nucleon $N_2$,  and
$f_{N_1}(z_1)$  is given by
\beq
f_{N_1}(z_1)\,=\,\int d{\bf k}_1 \,dE\,P_{N_1}(|{\bf k}_1| ,E)\,z_1\,\delta\left(z_1-\frac{k_1\cdot q}
{m_N\,\nu}\right)\,.
\label{effedue}
\eeq
where in  the quark-parton model,  the nucleon fragmentation structure functions
have the form $H_2^{N_1,N_2}=2xH_1^{N_1,N_2}$, with  $H_2^{N_1,N_2}$  given by Eq.~(\ref{h2n}).

\subsection{Results of calculations}
Taking into account the full FSI described by the
operator $\hat{S}_{FSI}$ of Eq.~(\ref{fsiop}), we have calculated
the differential cross section of the process $^{12}C(e,e^\prime p)X$ given by
\begin{equation}
\frac{d^4\sigma_{tf}}{dE_e^\prime d\Omega_e^\prime dT_2d\Omega_2}\,=\,{\widetilde
K(x,y,Q^2)}\,F_2^{(N_1/A,\,FSI)}(x,\ner{p}_2),
\label{crossfigure}
\end{equation}
where
\begin{equation}
\widetilde K(x,y,Q^2,T_2)= \frac{4\,\alpha^2\,E_e\,E_e^\prime}{\nu\,Q^4}\,(1-y+y^2)(T_2+m_N)
(T_2^2+2\,m_N\,T_2)^{1/2}
\label{coeff}
\end{equation}
 The results of our
calculations are presented in Figs.~\ref{Fig9}-\ref{Fig12}, where
the separate  contributions of the various kinds of  FSI and  their
summed effect are shown
 {\it vs.}  the kinetic energy of the detected proton.
In order to compare with the results
of Ref. \cite{ciosim2},  calculations have been performed assuming an incident electron energy of
$E_e=20\,\,GeV$ and an electron scattering angle  $\theta_e=15^o$,
with values of the Bjorken scaling variable equal to $x=0.2$ and
$0.6$; the proton emission angle has been fixed at the  values
$\theta_2 = 25^o$ (\textit{forward proton emission}) and $\theta_2= 140^o$
(\textit{backward proton emission}). It can be seen that
the most relevant contribution of the  FSI is due, both in forward and
backward nucleon emissions, to the rescattering of the hadronizing
quark with the $(A-2)$- nucleon system. In agreement with Ref. \cite{ciosim2},
 the effects of FSI between
the recoiling nucleon and $(A-2)$-nucleon system amounts to an
attenuation factor which, in the analyzed proton momentum
$|\bp_2|$ range, decreases the cross section up to  a factor of two; as expected, this contribution is more relevant for low
values of the momentum.
 We also checked the
sensitivity of the process upon the model
for the effective cross section $\sigma_{eff}(z,x,Q^2)$,  describing the interaction of the hadronizing quark with the
 spectator
 nucleon; to this end we calculated the cross section which includes
  the final state interaction between  the nucleon debris
 and the detected nucleon using the time dependent
 $\sigma_{eff}(z,x,Q^2)$  of Ref. ~\cite{ciokop}, adopted in this paper, and a constant
 cross section $\sigma_{eff}=60\,mb$, used in Ref. ~\cite{marksemi} in the description of proton backward production from the deuteron.
 The results, which are  presented in  Fig.~\ref{Fig11},   appear to appreciably
 depend upon the model of $\sigma_{eff}$; such
  a dependence however is very mild in the kinematics considered in Ref. ~\cite{marksemi},
  characterized by
 very low values of the momentum of the detected nucleon $(|{\bf p}|_2 \lesssim 0.1 GeV/c)$.
  Eventually,
we analyzed the role of the fragmentation mechanism: the results,
presented in Fig.~\ref{Fig12}, show  that, as in the deuteron
case, the target fragmentation mechanism contributes to nucleon
emission in the forward  direction and becomes appreciable only at
high values of $T_2$ ($T_2>600$ $MeV$). It should be noted that
such large kinetic energy are beyond of applicability of our
approach and that  in the region $50$ $MeV<T_2<250$ $MeV$, where
the use of a non relativistic spectral function is well grounded,
the effects of target fragmentation play only a minor role.
From the results we have exhibited it turns out that although FSI are very important, they
should not  hinder, in principle,  the extraction of the bound nucleon structure functions,
since the $x$-dependence of $\sigma_{eff}(z,x,Q^2)$ is very mild (cf. Ref. \cite{ciokop})
so that the $x$-dependence
of Eq.~(\ref{crossfsi}) is almost entirely governed by 
the DIS  nucleon structure function
$F_2(x/z_1)$. One can therefore consider the ratio
\begin{equation}
R(x,x',{\bf p}_2) = \frac{F_2^{(N_1/A,\,FSI)}(x,\ner{p}_2)}{F_2^{(N_1/A,\,FSI)}(x',\ner{p}_2)}
\label{ratio}
\end{equation}
which is the generalization to the FSI case of the quantity
suggested in Ref. \cite{ciosim2}. In case of the deuteron the ratio in PWIA simply
reduces to the
quantity
 $F_2^{N/D}(x/z_1)/F_2^{N/D}(x'/z_1)$, whereas for complex nuclei such a direct relation
  between
 Eq.~(\ref{ratio}) and the bound nucleon structure functions  cannot  be obtained
  due to the combined effects  of the nuclear  convolution and the FSI. Concerning the effects of the
  latter,
  it should be
  pointed out that they are produced by the effective cross section $\sigma_{eff}(z,x,Q^2)$
   which exhibits
  only a mild dependence upon $x$, so that  the x-dependence of Eq. (\ref{ratio})
  will be still governed by the
 the nucleon structure functions $F_2^{N/A}(x/z_1)$. We are currently investigating this point, as well as other
 possible ways to extract $F_2^{N/A}$ from the experimental data on complex nuclei; this would provide precious information on the A-dependence
 of possible medium modifications of  nucleon properties which, at the same time,
  would represent a valuable contribution
  to a final understanding of the elusive EMC effect.

\section{Summary and Conclusions}

 We have considered  proton production in   Semi Inclusive DIS
 processes $A(e, e^\prime p)X$ within the spectator and the target fragmentation mechanisms,
  taking all kind of FSI into account.
 A systematic study of this  process is of  great relevance in hadronic physics.
  As a matter of fact,
 in case of  a deuteron target detailed information on the DIS neutron structure
 function could in
  principle be obtained by
 performing experiments in the kinematical region where FSI are minimized
 (backward production and parallel kinematics);
  at the same time,
 if the experiment is performed when FSI are  maximized (perpendicular kinematics)
  the nonperturbative QCD phenomenon of
  hadronization could be investigated.
In case of complex nuclei SIDIS  could  also represent a tool  to investigate
 short-range correlations in nuclei, for
 the main source of  backward protons  originates in a complex nucleus from a correlated pair,
 moreover, SIDIS on complex nuclei might in principle serve to investigate the A-dependence of
 possible medium induced modification of the DIS nucleon structure function. However
 being these experiments performed on nuclear targets one always face the longstanding
 problem of a careful treatment of nuclear effects, like e.g. the short range behavior of the
nuclear wave function and the effects of the FSI, which is a prerequisite before drawing
conclusion about medium induced modifications of nucleon properties. In this respect, we
like to point out that so far, apart for few exceptions concerning the deuteron \cite{ckk, marksemi},
the problem  of the FSI has been overlooked, in particular as far as the interaction of the
hadronizing quark with the nuclear medium is concerned. For this reason,  in the present paper:
i) we have improved the treatment of the FSI in the deuteron case by using a time-dependent
 effective cross section $\sigma_{eff}(z,x,Q^2)$,  describing the interaction of the hadronizing quark with the
 spectator
 nucleon, featuring the proper $Q^2$ behavior, and ii) we have calculated the SIDIS
 cross section off complex nuclei taking all types of FSI into account, namely
 the rescattering  of the leading hadronizing quark  with the recoiling proton and with
 the residual
 $(A-2)$-nucleon system, which, apart from our preliminary results \cite{noi},
  have not been considered in previous investigations of SIDIS off complex nuclei.

The  main results we have obtained  can be summarized as follows:
\begin{enumerate}
\item
in SIDIS off the deuteron  FSI effects are   minimized in backward emission
and  maximized in  perpendicular kinematics. In the  former case
the bound nucleon structure function can  be investigated, whereas in the latter
case  information on QCD hadronization mechanisms can be obtained;
\item
in the case of complex nuclei the reinteraction of the hadronizing quark with the spectator
$(A-2)$-nucleon system
 appreciably attenuates the cross section, since  the
 survival probability of the $(A-2)$ nucleus is strongly reduced \cite{ciokop1};
for this reason,   some doubts can be cast as to the possibility
to perform SIDIS experiments of the type we have considered, where the underlying mechanism is almost
fully exclusive, being the unobserved $(A-2)$ nucleus in a well defined energy state.
   A more realistic case would be to consider
   a really semi-inclusive
  process by summing over all energy states of $(A-2)$-nucleon system, when the effects from
   FSI are
  expected to be much smaller.
  Calculations of this type are in progress and
  will be presented elsewhere \cite{noi};
\item
as in Ref. \cite{ciosim2}, we found that the interaction of the
recoiling proton with the $(A-2)$-nucleon system is relevant only
at low proton kinetic energies, leading to an overall small
attenuation of the cross section;
\item
in agreement with Ref. \cite{marksemi}, we found that in case of a deuteron target,
 FSI and target fragmentation mechanisms
 play a secondary role in
slow proton production in the backward hemisphere, which  is governed by
the spectator mechanism,
 provided
$T_p\,\lesssim\,0.3\,GeV$ ($|{\bf p}_2|\lesssim 0.8\,GeV/c$);
\item
both for the deuteron and complex nuclei we found that at the highest considered proton energies,
in the forward hemisphere and  partly also in the backward one, the effects from
target fragmentation and FSI become important. Thus slow proton production in SIDIS could be
 a sensitive tool to investigate
non perturbative QCD effects. In this connection
it has been suggested \cite{brooks} that higher sensitivity to nonperturbative current and
target fragmentation mechanisms could be
achieved by detecting, in coincidence with the slow proton, the fast leading hadron arising
from current fragmentation. The extension of our approach to this process,
 which can experimentally be
 investigated by the CLAS detector at JLab, is straightforward;
 \item we did not address here in details the problem concerning the most
 reliable way of extracting
 from the experimental data on nuclei information on the DIS nucleon structure function but pointed
 out that the important role played by FSI should not in principle hinder such a possibility.
\end{enumerate}

In summary, slow hadron production in SIDIS appears to be a powerful tool to investigate
both the properties of bound nucleons and the  hadronization mechanisms.

\acknowledgments{This work is supported in part by the Italian Ministry of
 Research and University through the Program Rientro dei Cervelli.
 L.P.K. is indebted to  the
 University of Perugia and INFN, Sezione di Perugia, for warm hospitality
 and financial support. Useful
 discussions with W. Brooks, B. Kopeliovich and S. Kuhn and M. Strikman
 are gratefully acknowledged.
 M.A. is supported by DOE grant
 under contract DE-FG02-93ER40771}.

%
\newpage
\begin{figure}[!hc]      
\centerline{\hspace{-.1cm}
\includegraphics[scale=1.0,angle=0]{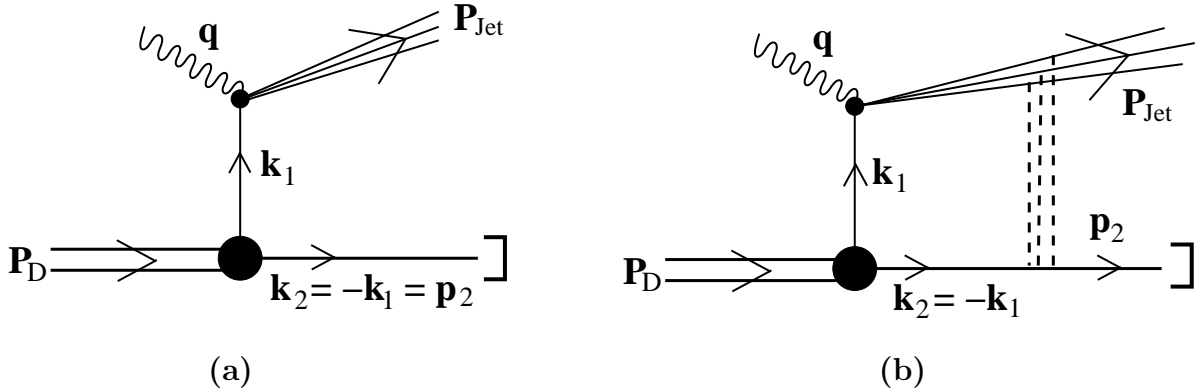}}
\caption{The Feynman diagrams of the process $D(e,e^\prime p)X$
  within: a) the spectator mechanism in PWIA and b) taking
  into account  FSI. ${\bf k}_1$ and
  ${\bf k}_2 = -{\bf k}_1$ are the nucleon three-momenta in the
  deuteron before $\gamma^*$  absorption and ${\bf p}_2$
  is the three-momentum of the detected proton~$p$.}
  \label{Fig2}
\end{figure}
\newpage
\begin{figure}[!ht]      
\centerline{
\includegraphics[scale=0.4 ,angle=0]{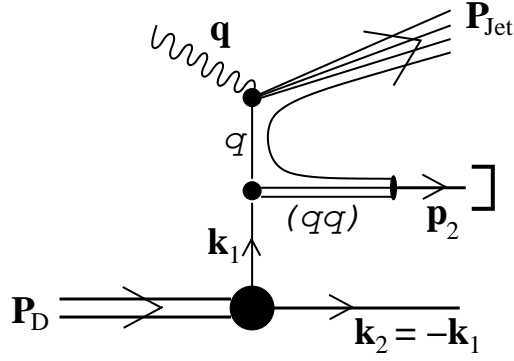}}
\caption{Proton production by  target fragmentation in the
  $D(e,e^\prime p)X$ process. The diquark $({q\,q})$ captures
  a quark  from the vacuum and the proton $p$ is formed
  and detected with three-momentum~${\bf p}_2$.}
  \label{Fig3}
\end{figure}
\vskip 3.0cm

\begin{figure}[!ht]      
\centerline{
\includegraphics[scale=0.55 ,angle=0]{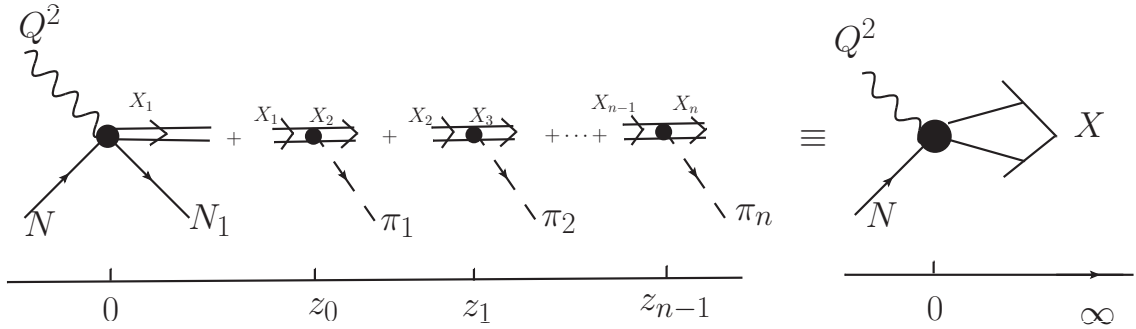}}
\caption{Schematic representation of pion production by quark  hadronization.}
\label{Fig4}
\end{figure}
\begin{figure}[!hb]      
\includegraphics[scale=0.8,angle=0]{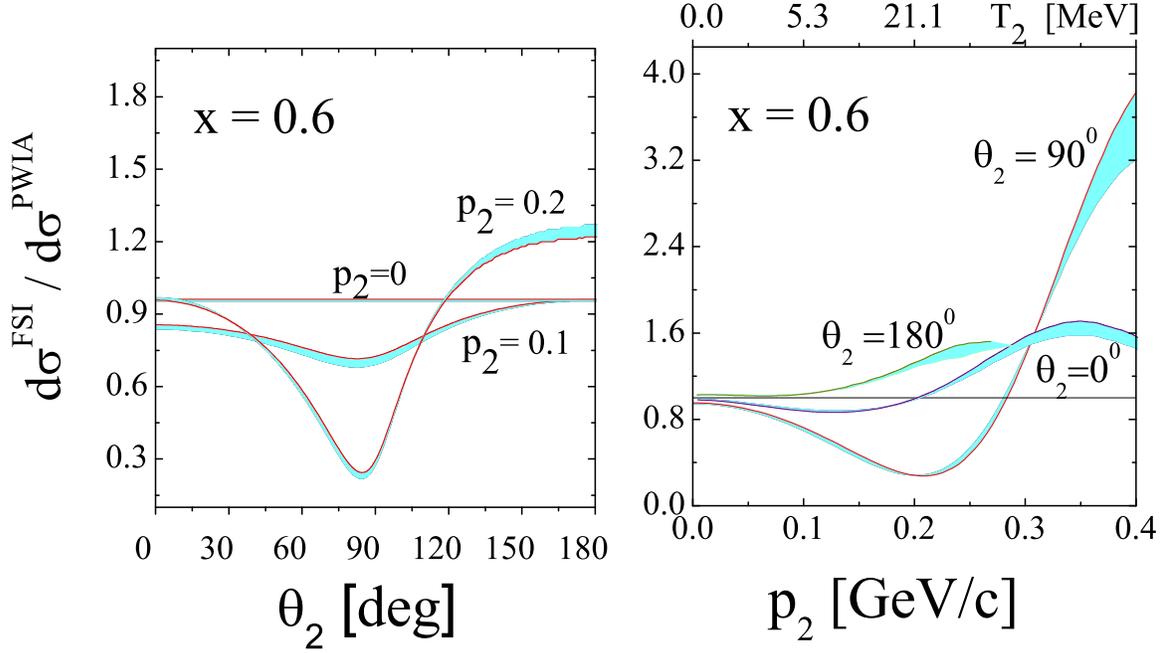}
\caption{The role of the FSI in the process $D(e,e^\prime p)X$
  within the spectator mechanism. {\it Left panel}: angular
  dependence of the ratio of the  cross section which includes
  FSI (Eq. (\ref{crossfsi})) to the PWIA cross section (Eq.
  (\ref{crossPWIA})), at several fixed values of the detected
  proton momentum $|\bp_2|\equiv p_2$ (in GeV/c). {\it Right panel}:
  dependence of the same ratio upon $p_2$ at parallel ($\theta=0^o$
  and $\theta=180^o$) and perpendicular ($\theta=90^o$) kinematics.
  Calculations have been performed at $Q^2=12\,  (GeV/c)^2$. The chosen
  kinematics is close to the one  planned in the future experiments
  at JLAB at  $E_e \sim 10 \,GeV$. The shaded area is due to
  the uncertainties in the  parameters appearing in Eq. (\ref{gamma}) (see Ref. \cite{ckk}).
  }\label{Fig5}
\end{figure}
\begin{figure}[!ht]      
\includegraphics[scale=0.8,angle=0]{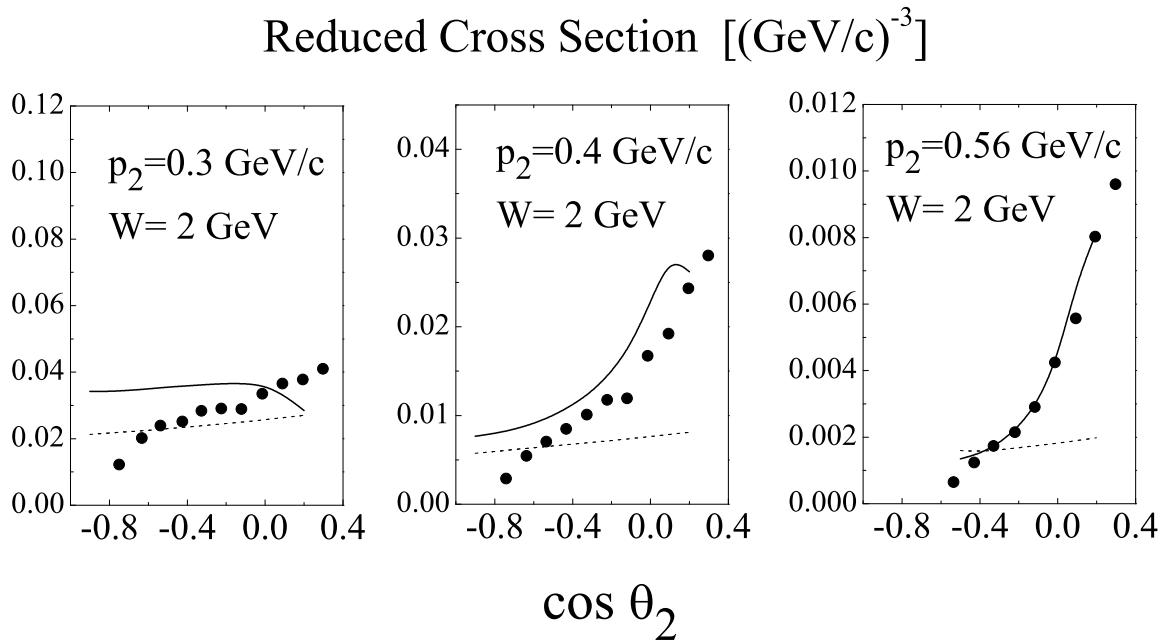}
\caption{The reduced cross section (full dots), i.e.
the  experimental cross
  section divided by the  kinematical factor $K(x, y, Q^2)$ (Eq.(\ref{kinfac}))
   \cite{Klimenko},
  {\it  vs.} the proton emission angle (the angle between {\bf q} and ${\bf p}_2$),
  at various
  values of $|\bp_2|$ and   fixed values of the four-momentum
  transfer ($Q^2=1.8\ (GeV/c)^2$) and  the invariant mass of the
  debris $X$, $W_X =\sqrt{(P_D-p_2+q)^2}\simeq W$. The dotted curve represents
  the   PWIA cross section ( Eq.(\ref{crossPWIA})) divided by the
  kinematical factor $K(x, y, Q^2)$, whereas the full curve represents the cross section
  (Eq.(\ref{crossfsi}))
   which includes the FSI
  between the hadronizing quark and the spectator nucleon,
  divided by the same kinematical factor  $K(x, y, Q^2)$. Note that within the PWIA
   the reduced cross
  section represents the product of the neutron DIS structure
  function $F_{2}^n(x/z_1,Q^2)$ and the deuteron momentum distribution
  $n_D(|\bp_2|)$ (Eq.(\ref{dismom})); since the latter does not
  depend upon the angle $\theta_2$, the angle dependence is only
  given by the quantity $x/z_1$, which is almost constant in the
  considered set of data. The inclusion of the FSI generates a strong  $\theta_2$
   dependence of the distorted momentum distributions $n_D^{FSI}({\bf q},{\bf p}_2)$
    (Eq. (\ref{dismomfsi})), with  the role of the FSI increasing with the value of $|\bp_2|$ due to the rapid
    fall off of the undistorted momentum distribution.}
  \label{Fig6}
\end{figure}
\begin{figure}[!ht]      
\centerline{
  \includegraphics[scale=0.75]{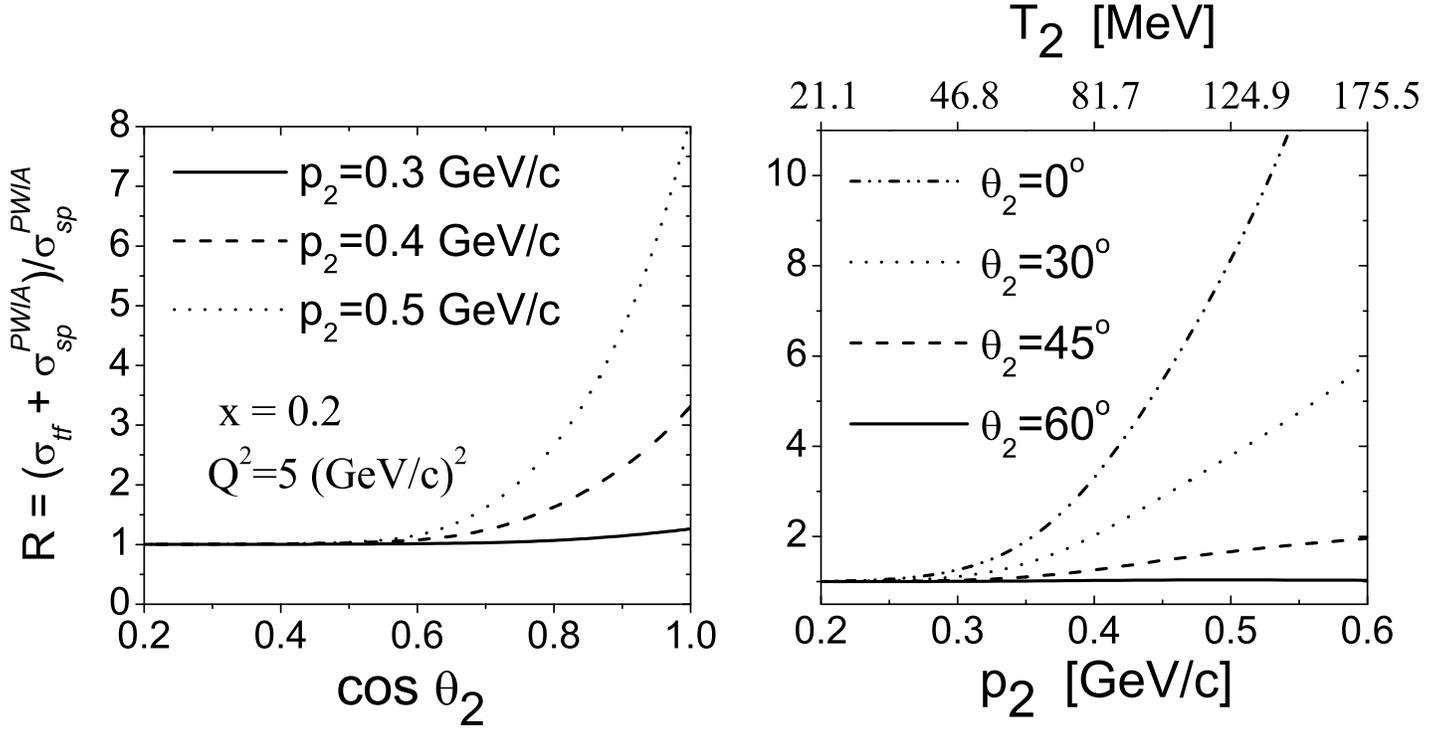}}
\caption{Contribution of  target fragmentation to nucleon emission
  in the process $D(e,e^\prime p)X$. The ratio
  of the sum of the cross sections (\ref{crossPWIA}) and
  (\ref{tf}) to the cross section (\ref{crossPWIA}),
  $R_{tf}=(d\sigma_{tf}+d\sigma_{sm}^{PWIA})/d\sigma_{sm}^{PWIA}$,
  plotted {\it  vs.} $\cos\theta_2$ and  {\it  vs.} $|\bp_2|\equiv p_2$ are shown in the left
   and right Figures, respectively. For convenience  the corresponding
  values of the kinetic energy $T_2$ of the proton is also displayed on the upper axis of
  right panel.}
\label{Fig7}
\end{figure}
\begin{figure}[!ht]      
\begin{center}
\includegraphics[width=4.8cm,height=3.9cm]{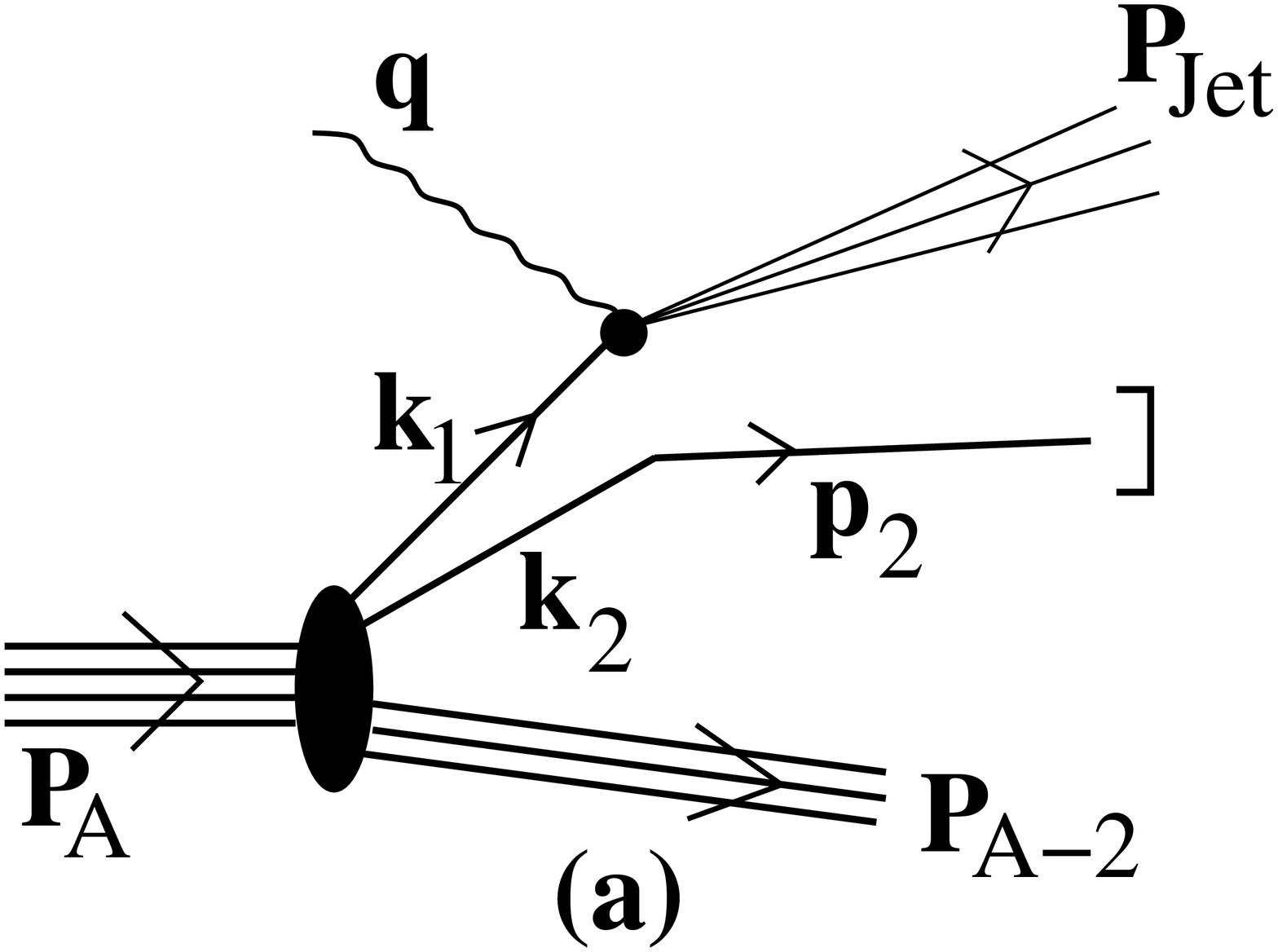}
\hspace{0.2cm}
\includegraphics[width=4.8cm,height=3.7cm]{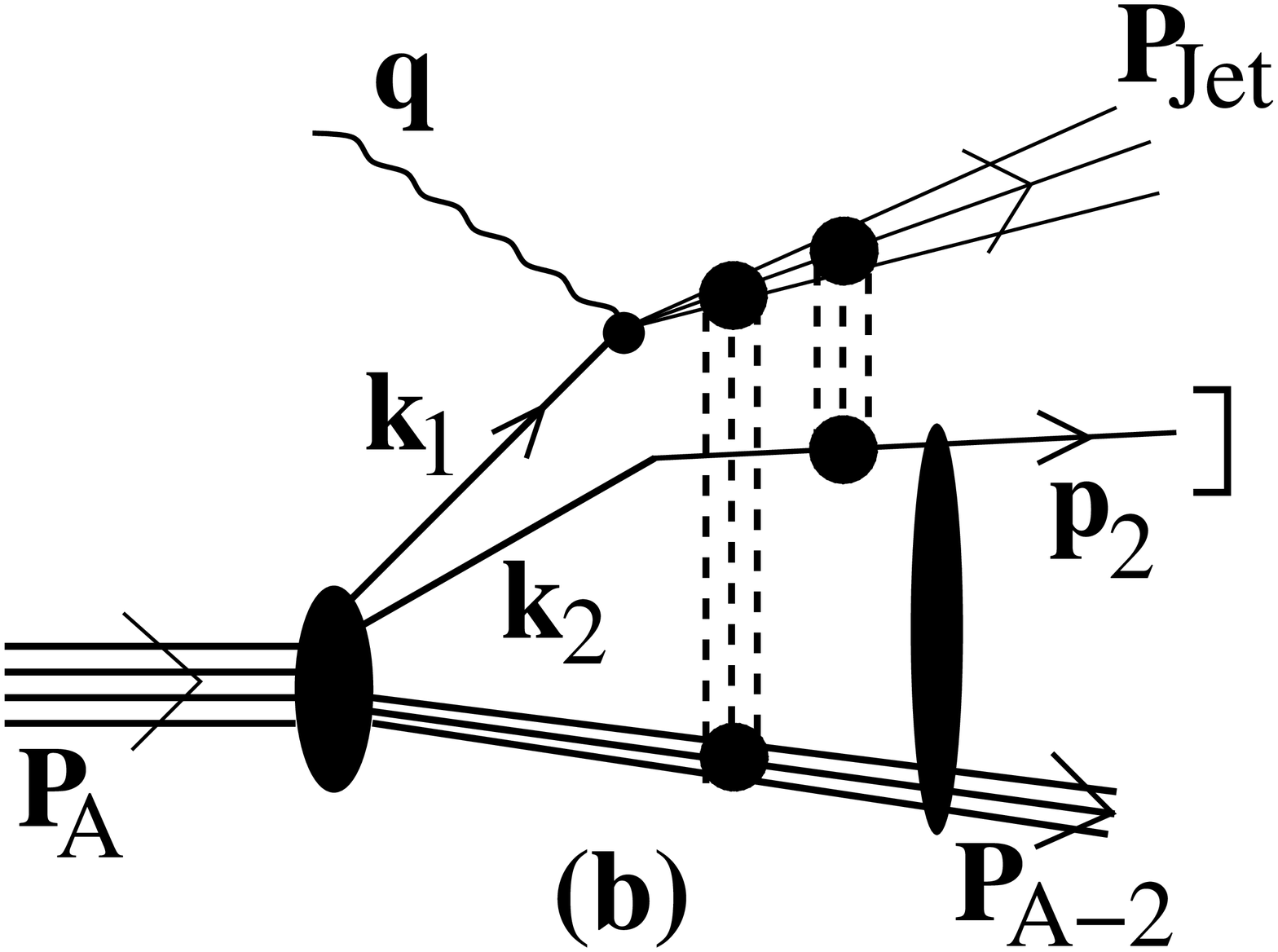}
\hspace{0.2cm}
\includegraphics[width=4.8cm,height=3.5cm]{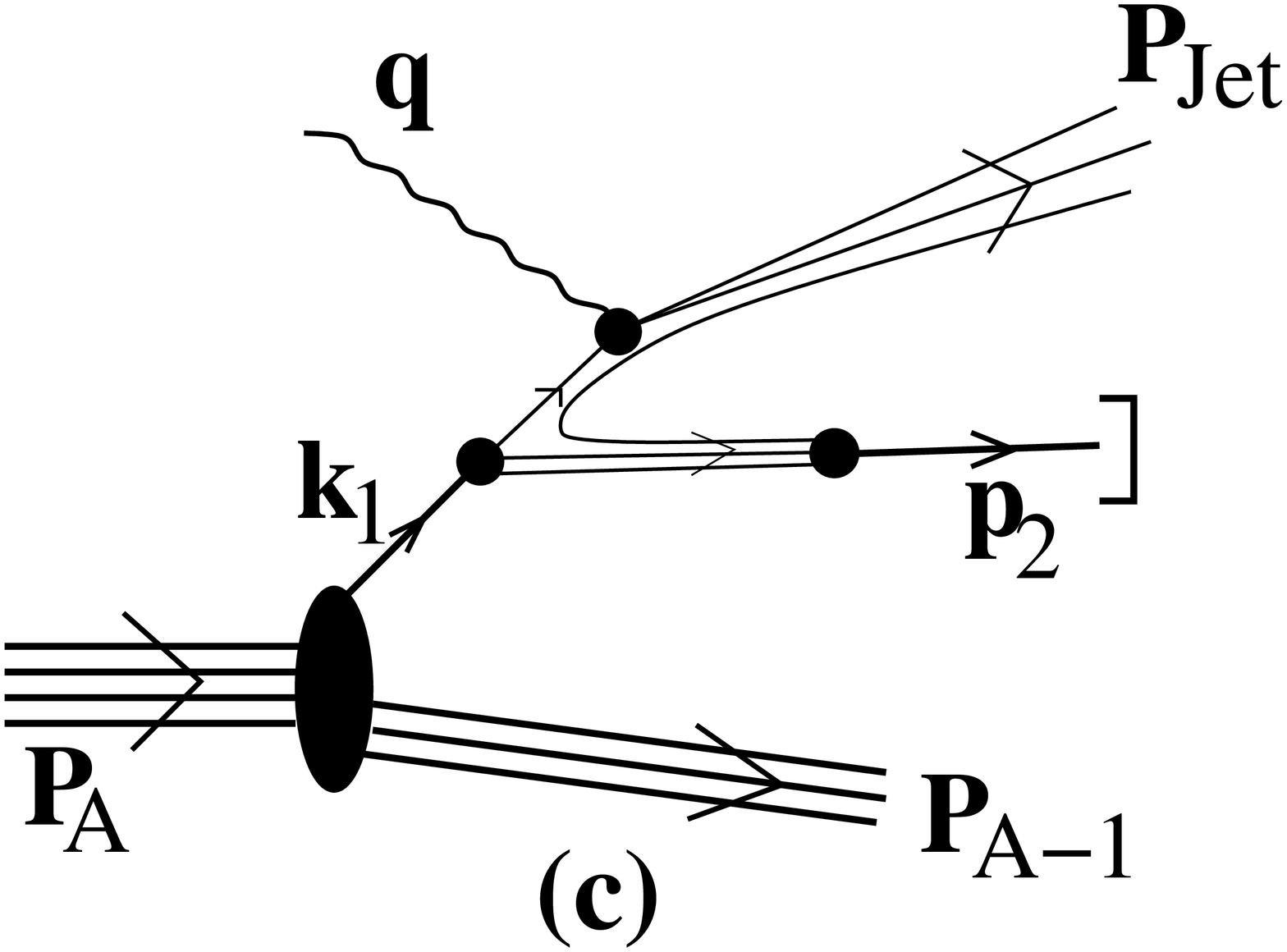}
\vskip -0.1cm \caption{Proton production in   $A(e,e^\prime p)X$
processes off a complex nucleus $A$: (a) spectator mechanism
within the PWIA; (b) various contribution to the FSI within the
spectator mechanism;  (c)  proton production from target
fragmentation. In each of the three processes a proton  with momentum ${\bf p}_2$, formed by
different mechanisms, is detected in
coincidence with the scattered electron.}
\label{Fig8}
\end{center}
\end{figure}
\begin{figure}[!ht]   
\centerline{
  \includegraphics[scale=0.75]{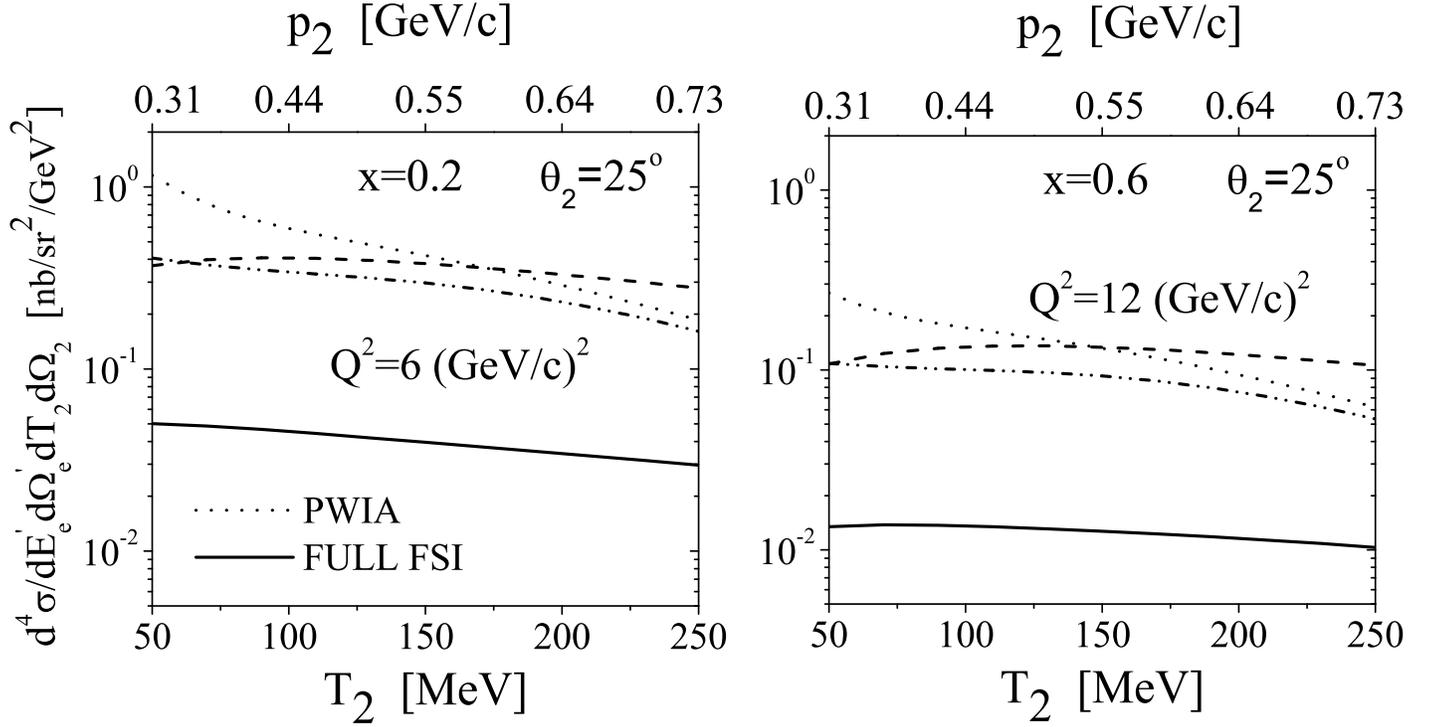}}
\caption{The SIDIS differential cross section for the process
  $^{12}C(e,e^\prime p)X$ {\it vs} the  kinetic energy $T_2$ of the detected
  proton, emitted forward at $\theta_2=25^0$,  in
  correspondence of two values of the Bjorken scaling variable $x$.
  {\it Dotted curve}: PWIA (Fig. 7(a));  {\it Dashed curve}:
  PWIA plus  FSI of the nucleon debris $X$ with the recoiling proton;
  {\it Dashed-double-dotted  curve}: PWIA plus  FSI of the proton with
  $(A-2)$-nucleon system; {\it Full curve}: PWIA plus
  the full FSI (Fig. 7(b)). For the sake of  convenience, on the upper axis the corresponding
values of the proton momentum $|{\bf p}_2|$ are also displayed.}

  \label{Fig9}
\end{figure}
\begin{figure}[!hb]           
\centerline{
  \includegraphics[scale=0.75]{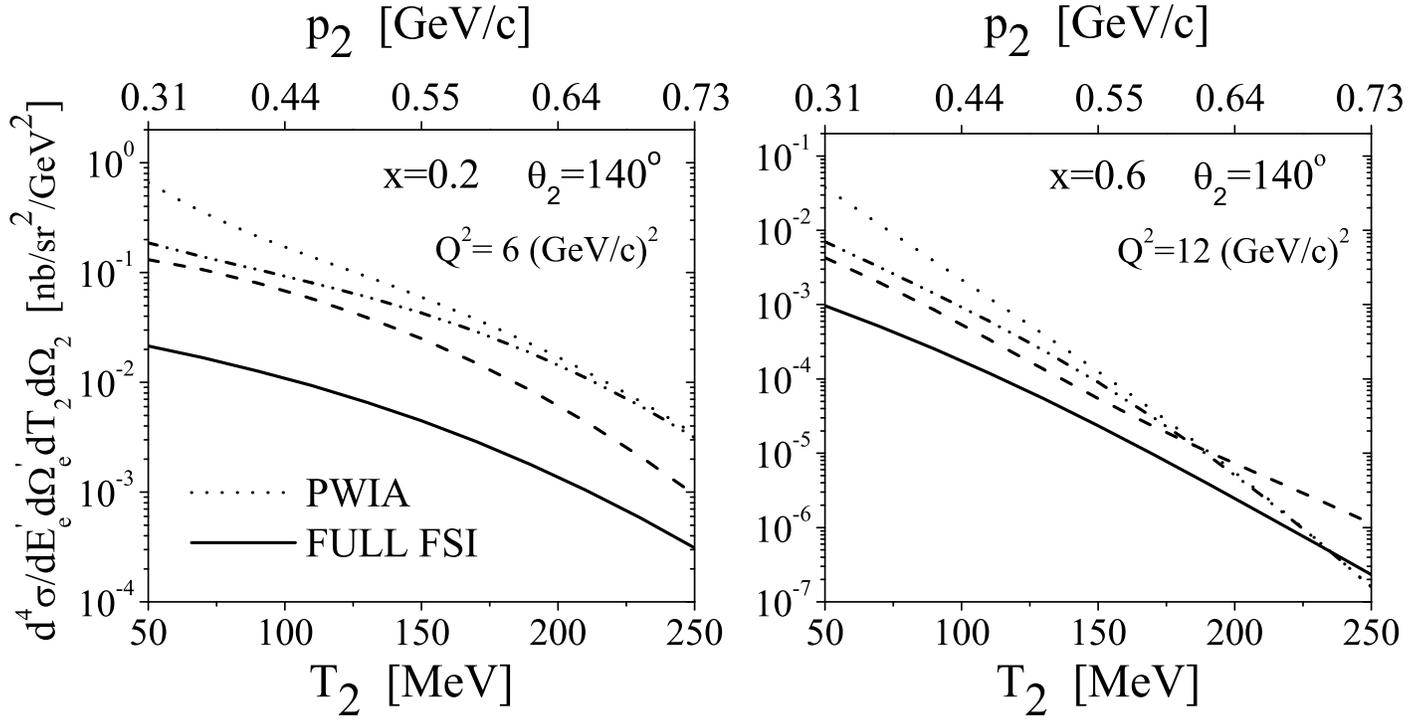}}
\caption{The same as in Fig. \ref{Fig9} for  protons
  emitted backward at $\theta_2=140^0$. }
  \label{Fig10}
\end{figure}
\begin{figure}[!hc]         
\centerline{
  \includegraphics[scale=0.75]{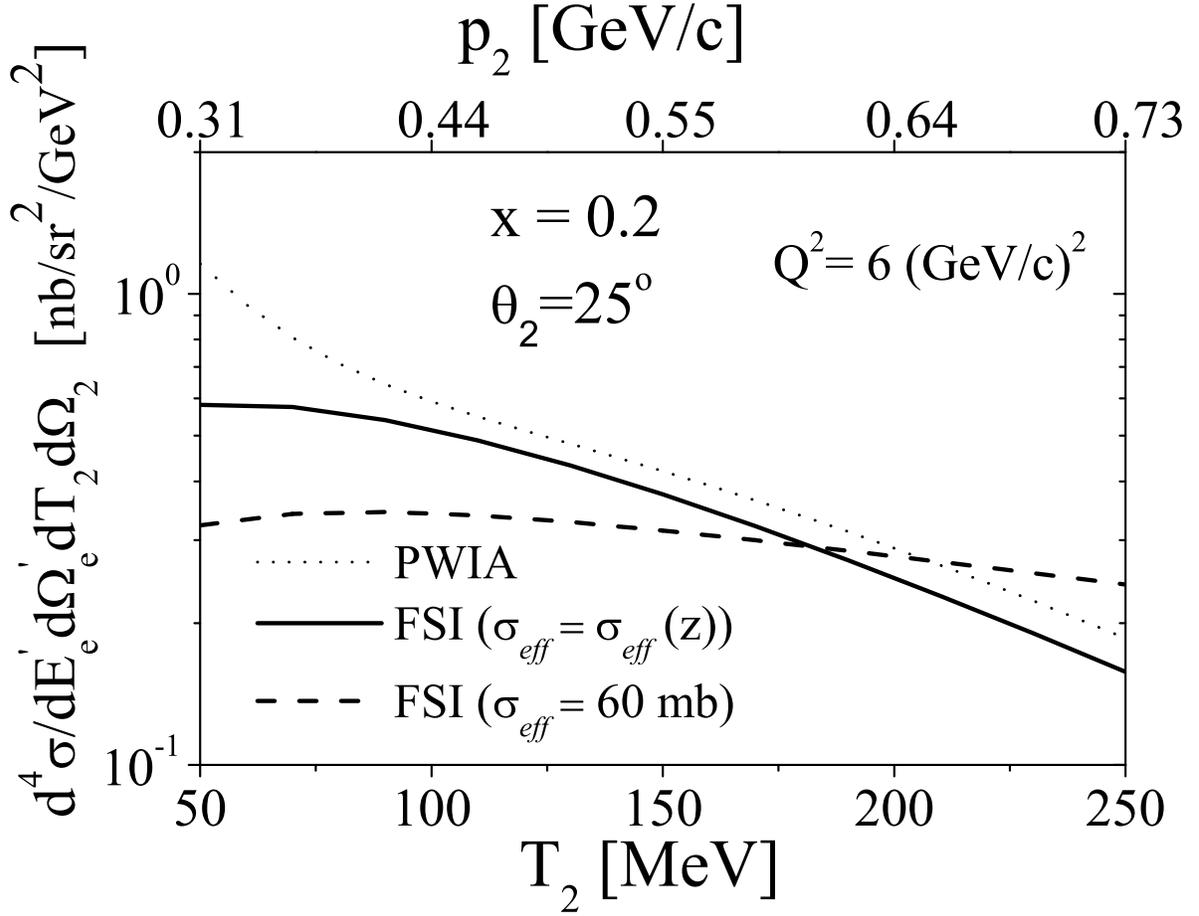}}
\caption{The SIDIS differential cross section for the process
  $^{12}C(e,e^\prime p)X$ with the FSI between the nucleon
  debris and the spectator nucleon  calculated at forward kinematics with the time
  dependent $\sigma_{eff}=\sigma_{eff}(z)$ \cite{ciokop} ({\it dashed curve})
  and with a constant $\sigma_{eff}=60\,\,mb$ ({\it full curve}). The
  PWIA results are presented by the {\it dotted curve}.}
  \label{Fig11}
\end{figure}
\begin{figure}[!ht]      
\centerline{
  \includegraphics[scale=0.75]{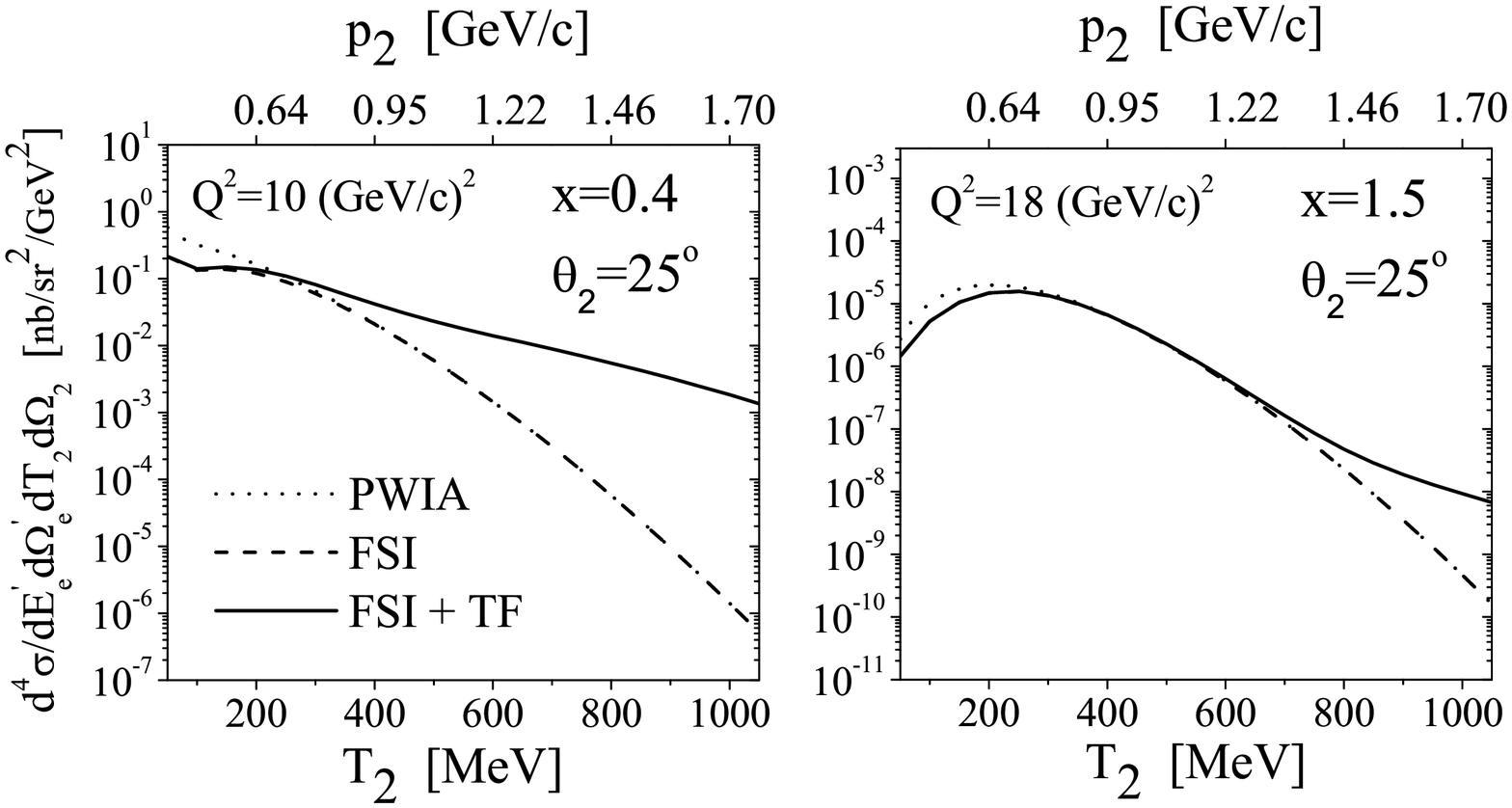}}
\caption{Proton production by target fragmentation in  the process
  $^{12}C(e,e^\prime p)X$  {\it vs} the  kinetic energy $T_2$ of the detected
  proton, emitted forward at $\theta_2=25^0$ at $x=0.4$ and
  $x=1.5$. {\it Dotted curve}: spectator mechanism within the PWIA; {\it dashed
  curve}: spectator mechanism within the PWIA plus FSI of the
  spectator nucleon with the $(A-2)$-nucleon system; {\it dot-dashed
  curve}: spectator mechanism within the PWIA plus FSI of the
  spectator nucleon with $(A-2)$-nucleon system plus target
  fragmentation. Note the different kinetic energy range in this and the previous figures.}
  \label{Fig12}
\end{figure}


\begin{thebibliography}{15}
%
\bibitem{FS}        L. L. Frankfurt and M. I. Strikman, Phys. Rep. {\bf 76} (1981) 216; {\it ibidem} {\bf 160} (1988) 235.
%
\bibitem{bosveld}    G. D. Bosveld, A. E. L. Dieperink, O. Scholten,
                     Phys. Rev. {\bf C49} (1994) 2379.
%
\bibitem{ciosim2}   C. Ciofi degli Atti and S. Simula, Phys. Lett. {\bf B319} (1993) 23;
                    C. Ciofi degli Atti and S. Simula, Few-Body Systems  {\bf 18} (1995) 55.
%
\bibitem{simula}    S. Simula, Phys. Lett. {\bf B387} (1996) 245.
%
\bibitem{sarg}      W. Melnitchouk, M. Sargsian and  M. I. Strikman, Z. Phys. {\bf A359} (1997) 99.
%

\bibitem{scopetta}  C. Ciofi degli Atti, L. P. Kaptari and S. Scopetta,
Eur. Phys. J. {\bf A5} (1999) 191.
%
\bibitem{sarg1}     M.M. Sargsian, J. Arrington, W. Bertozzi, W. Boeglin
 et al., J. Phys. {\bf G29} (2003) R1.
%
\bibitem{ckk}       C. Ciofi degli Atti, L. P. Kaptari, B.Z. Kopeliovich, Eur.
 Phys. J. {\bf A19} (2004) 145.
 %
 \bibitem{marksemi} M. Sargsian, M. Strikman, Phys. Lett. {\bf B639} (2006)
 223.
 %
\bibitem{experiment}E. Matsinos, et al., Z. Phys. {\bf C44} (1989) 79;\\
                    T. Kitagaki, et al., Phys. Lett. {\bf B214} (1988) 281;\\
                    G. Guy et al, Phys. Lett. {\bf B229} (1989) 421;\\
                    M. R. Adams et al, Phys. Lett. {\bf B319} (1993) 23.

\bibitem{Klimenko}  A.V. Klimenko, S.E. Kuhn, C. Butuceanu, K.S. Egiyan et al.,
 Phys. Rev. {\bf C73} (2006) 035212;\\
 S.E. Kuhn, {\it Private communications}.
%
%

%
\bibitem{wally} H. Fenker, C. Keppel, S. Kuhn, and W. Melnitchouk (spokespersons), {\it The Structure
of the Free Neutron Via Spectator Tagging}, JLab proposal E-03-012 (2003).
%
\bibitem{brooks}    W. Brooks and H. Hakobyan, in \textit{Sixth International
Conference on Perspectives in Hadronic Physics}, S. Boffi, C.
Ciofi degli Atti, M. Giannini, D. Treleani Eds., AIP Conference
Proceedings, Vol.1056 (2008) 215.
%
\bibitem{leading}   A. Airapetian,   et. al, HERMES Coll., Nucl. Phys. {\bf B780} (2007) 1.
%
\bibitem{ciokop}    C. Ciofi degli Atti and B. Kopeliovich, Eur. Phys.J. {\bf A17} (2003) 133.
%
\bibitem{ciokop1}   C. Ciofi degli Atti and B. Kopeliovich, Phys. Lett. {\bf B606} (2005) 281.
%
\bibitem{adams}     E665 Collaboration, M. R. Adams {\it et al} Z. Phys. {\bf C65} (1995) 225.
%
\bibitem{Ffunction} L. Trentadue and G. Veneziano,  Phys. Lett. {\bf B323} (1994) 201.
%
\bibitem{distrtr}   S. L. Wu, Phys. Rep. \textbf{107} (1984) 59.
%
\bibitem{barframaj} A. Bartl, H. Fraas, W. Majerotto, Phys. Rev.  \textbf{D26} (1982) 1061.
%
\bibitem{nashtobe}  M. Alvioli, C. Ciofi degli Atti, L.P. Kaptari, C. B. Mezzetti, V. Palli
 {\it to be published}.
%
\bibitem{ciosim}    C. Ciofi degli Atti and S. Simula, Phys. Rev. {\bf C53} (1996) 1689.
%
\bibitem{noi}       M. Alvioli, C. Ciofi degli Atti and V. Palli, Nucl. Phys.A \textbf{782} (2007) 175c.
%
\bibitem{glau1}     R. J. Glauber, Phys. Rev. \textbf{100} (1955) 242.
%
\bibitem{glau2}     R. J. Glauber, High-energy Collision Theory, in W. E. Brittin and L. Dunham editors,
                    \emph{Lectures in Theoretical physics} , Ed. W. Brittin, N. Y. Interscience 1959;\\
                    R. J. Glauber, ``\emph{High Energy Physics and Nuclear Structure}'',
                    Ed. G.Alexander, North Holland, 1967; Ed. S. Devons, Plenum Press,1970.
%
\bibitem{arn}       R. A. Arndt et al. ``(SAID) partial wave analisys facility'',
                    http://said.phys.vt.edu.
%
\bibitem{koppred}   B. Z. Kopeliovich, J. Nemchik, E. Predazzi and A. Hayashigaki, Eur. Phys. J. A
                    \textbf{19S1} (2004) 111.
%
\bibitem{alvmarpal} M. Alvioli, C. Ciofi degli Atti, I. Marchino, C. Mezzetti and V. Palli,
{\it To appear}.
\end{thebibliography}
\end{document}